\documentclass[aps,prd,nofootinbib,showkeys,preprint,floatfix]{revtex4-1}

\pdfoutput=1

\usepackage{graphicx}
\usepackage{graphics}
\usepackage{tikz}
\usepackage{makecell}
\usepackage[mathscr]{eucal}
\usepackage{amssymb}
\usepackage{amsmath}
\usepackage{xspace}
\usepackage{listings}
\usepackage{ulem}
\usepackage{color}
\usepackage{alltt}

\definecolor{darkgreen}{rgb}{0,0.5,0}

\newcommand{\vev}[0]{VEV\xspace}
\newcommand{\vevs}[0]{VEVs\xspace}
\newcommand{\DRbar}[0]{{\ensuremath{\overline{\mathrm{DR}}}}\xspace}
\newcommand{\drbarp}[0]{{\ensuremath{{\overline{\mathrm{DR}}}'}}\xspace}
\newcommand{\msbar}[0]{{\ensuremath{{\overline{\mathrm{MS}}}}}\xspace}
\newcommand{\gev}[0]{\text{ GeV}\xspace}

\newcommand{\eq}[0]{eq.\xspace}
\newcommand{\eqs}[0]{eqs.\xspace}

\newcommand{\fig}[0]{Fig.\xspace}

\newcommand{\eg}[0]{\textit{e.g.}\xspace}

\newcommand{\ie}[0]{\textit{i.e.}\xspace}

\newcommand{\etc}[0]{\textit{etc.}\xspace}

\newcommand{\Sec}[0]{sec.\xspace}
\newcommand{\App}[0]{app.\xspace}
\def\gsim{\raise0.3ex\hbox{$\;>$\kern-0.75em\raise-1.1ex\hbox{$\sim\;$}}}
\newcommand{\cpp}[0]{\texttt{C++}\xspace}
\newcommand{\py}[0]{\texttt{Python}\xspace}

\newcommand{\sarah}[0]{\texttt{SARAH}\xspace}
\newcommand{\mathematica}[0]{\texttt{Mathematica}\xspace}
\newcommand{\vevacious}[0]{\texttt{Vevacious}\xspace}
\newcommand{\vcs}[0]{\vevacious}
\newcommand{\homps}[0]{\texttt{\texttt{HOM4PS2}}\xspace}
\newcommand{\lhpc}[0]{\texttt{\texttt{LHPC}}\xspace}
\newcommand{\pyminuit}[0]{\texttt{PyMinuit}\xspace}
\newcommand{\pym}[0]{\pyminuit}
\newcommand{\minuit}[0]{\texttt{MINUIT}\xspace}
\newcommand{\cosmotransitions}[0]{\texttt{CosmoTransitions}\xspace}
\newcommand{\ct}[0]{\cosmotransitions}
\newcommand{\spheno}[0]{\texttt{SPheno}\xspace}
\newcommand{\xml}[0]{\texttt{XML}\xspace}
\newcommand{\xmle}[1]{\texttt{$<$#1$>$}\xspace}
\newcommand{\slha}[0]{\texttt{SLHA}\xspace}
\newcommand{\block}[0]{\texttt{BLOCK}\xspace}
\newcommand{\blocks}[0]{\texttt{BLOCK}s\xspace}

\newcommand{\softsusy}[0]{\texttt{SoftSUSY}\xspace}
\newcommand{\suspect}[0]{\texttt{SuSpect}\xspace}

\newcommand{\isajet}[0]{\texttt{ISAJET}\xspace}

\newcommand{\AddrBonn}{%
Bethe Center for Theoretical Physics \& Physikalisches Institut der
 Universit\"at Bonn, \\
53115 Bonn, Germany }

\newcommand{\AddrWur}{%
Institut f\"ur Theoretische Physik und Astronomie,
Universit\"at W\"urzburg\\
Am Hubland,
97074 Wuerzburg}

\begin{document}

\title{\vcs: A Tool For Finding The Global Minima Of One-Loop Effective
       Potentials With Many Scalars}

\author{J.\ E.\ Camargo-Molina} \email{jose.camargo@physik.uni-wuerzburg.de}

\author{B.\ O'Leary} \email{ben.oleary@physik.uni-wuerzburg.de}

\author{W.\ Porod} \email{porod@physik.uni-wuerzburg.de}\affiliation{\AddrWur}

\author{F.\ Staub}\email{fnstaub@th.physik.uni-bonn.de}
\affiliation{\AddrBonn}


\pacs{??, ??, ??}

\preprint{Bonn-TH-2013-08}
\begin{abstract}
Several extensions of the Standard Model of particle physics contain additional
 scalars implying a more complex scalar potential compared to that of the
 Standard Model. In general these potentials allow for charge- and/or
 color-breaking minima besides the desired one with correctly broken
 $SU(2)_{L} \times U(1)_{Y}$.  Even if one assumes that a metastable local
 minimum is realized, one has to ensure that its lifetime exceeds that of our
 universe. We introduce a new program called \vcs which takes a generic
 expression for a one-loop effective potential energy function and finds
 {\em all} the tree-level extrema, which are then used as the starting points
 for gradient-based minimization of the one-loop effective potential. The
 tunneling time from a given input vacuum to the deepest minimum, if different
 from the input vacuum, can be calculated. The parameter points are given as
 files in the \slha format (though is not restricted to supersymmetric models),
 and new model files can be easily generated automatically by the \mathematica
 package \sarah. This code uses \homps to find all the minima of the tree-level
 potential, \pym to follow gradients to the minima of the one-loop potential,
 and \ct to calculate tunneling times.
\end{abstract}

\maketitle


\section{Introduction}
\label{sec:introduction}

A major part of the phenomenology of the incredibly successful standard model of
 particle physics (SM) is the spontaneous breaking of some (but not all) of the
 gauge symmetries of the Lagrangian density by the
 \textit{vacuum expectation value} (\vev) of a scalar field charged under a 
 subgroup of the SM gauge group. The entire scalar sector of the SM consists of
 a doublet of $SU(2)_{L}$ which also has a hypercharge under $U(1)_{Y}$ equal
 in magnitude to that of the lepton $SU(2)_{L}$ doublet. The potential energy of
 the vacuum is minimized by the scalar field taking a constant non-zero value
 everywhere. The presence of this \vev radically changes the phenomenology of
 the theory, and allows for masses for particles that would be forced to be
 massless if the gauge symmetries of the Lagrangian density were also symmetries
 of the vacuum state.

Since this scalar field is the only field in the SM that can possibly have a
 non-zero \vev while preserving Lorentz invariance, finding the minima of the
 potential energy is straightforward, though of course evaluating it to the
 accuracy required is quite involved
 \cite{EliasMiro:2011aa, Bezrukov:2012sa, Degrassi:2012ry}.

Also, with the current measurements for the masses of the top quark and
 Higgs boson, one finds that the SM potential at one-loop order is actually
 unbounded from below for a fixed value of the renormalization scale.
 The value of the Higgs field for which the potential is
 lower than the desired vacuum is so high that one may worry that large
 logarithms of the Higgs field over the electroweak scale would render the loop
 expansion non-convergent. However, the effect of large logarithms can be
 resummed, and the conclusion that our vacuum is only metastable persists using
 the renormalization-group-improved effective potential
 \cite{Isidori:2001bm, Ellis:2009tp, EliasMiro:2011aa, Bezrukov:2012sa,
 Degrassi:2012ry}.

The existence of multiple non-equivalent vacua both raises technical challenges
 and introduces interesting physics. The technical challenges are now that one
 has to find several minima and evaluate which is the deepest, as well as
 calculate the tunneling time from a false vacuum to the true vacuum. However,
 this is an important ingredient in theories where a first-order phase
 transition explains the baryon asymmetry of the universe through the sphalerons
 occuring in the nucleation of bubbles of true vacuum (see \cite{Riotto:1999yt}
 and references therein).

Many extensions of the SM introduce extra scalar fields. Sometimes these fields
 are introduced explicitly to spontaneously break an extended gauge symmetry
 down to the SM gauge group \cite{Langacker:2008yv, Basso:2010si}, and they are
 assumed to have non-zero \vevs at the true vacuum of the theory. Other times
 they are introduced for other reasons, such as supersymmetry
 \cite{Nilles:1983ge}, and often non-zero \vevs for such fields would be
 disastrous, such as breaking $SU(3)_{c}$ and/or $U(1)_{\text{EM}}$, which
 excludes certain parts of the parameter space of the minimal
 supersymmetric standard model (MSSM) from being phenomenologically relevant.

The technical challenges are much tougher when multiple scalar fields are
 involved. Even a tree-level analysis involves solving a set of coupled cubic
 equations, the so-called minimization or tadpole equations. It has generally
 only been attempted for highly symmetric systems such as two Higgs doublet
 models (2HDM) \cite{Lee:1973iz, Branco:2011iw} or with only a minimal amount of
 extra degrees of freedom such as the (assumed) three non-zero \vevs of the
 next-to-minimal supersymmetric standard model (NMSSM)
 \cite{Fayet:1974pd, Ellis:1988er, Drees:1988fc}.

Since a general solution is usually too difficult, the question of the
 stability of \vev configurations against tunneling to other minima of the
 potential is often ignored. Instead, potentials are often engineered to have a
 local minimum at a desired \vev configuration through ensuring that the tadpole
 equations are satisfied for this set of \vevs. This approach allows one to go
 beyond tree level straightforwardly, and one-loop tadpoles are the norm, and
 in supersymmetric models two-loop contributions are often included
 \cite{Dedes:2002dy}. This local minimum is implicitly assumed to be stable or
 long-lived enough to be physically relevant. Unfortunately, as some examples
 will show, local minima which are not the global minimum of their parameter
 point are often extremely short-lived, excluding some benchmark parameter
 points for some models.

The program \vcs has been written to address this. Given a set of tadpole
 equations and the terms needed to construct the one-loop effective
 potential\footnote{The potential may include non-renormalizable terms, as
 long as the one-loop effective potential is of the form of a polynomial plus
 $V^{\text{mass}}$ as given in \eq~\ref{eq:potential_loop_corrections}.},
 first all the extrema of the tree-level potential are found using homotopy
 continuation (\homps), which are then used as starting points for
 gradient-based minimization (\pyminuit) of the (real part of the) one-loop
 potential, and finally, if requested, the tunneling time from an input minimum
 to the deepest minimum found is estimated at the one-loop level
 (\cosmotransitions). The program is intended to be suitable for parameter
 scans, taking parameter points in the \slha format
 \cite{Skands:2003cj, Allanach:2008qq} and giving a result within seconds,
 depending on the number of fields allowed to have non-zero \vevs and the
 accuracy of the tunneling time required.

\begin{center}
 \vcs is available to download from\\
 {\tt http://www.hepforge.org/downloads/vevacious}.
\end{center}

\section{The potential energy function at tree level and one-loop level}
\label{sec:potential}

The terminology of minimizing the effective potential of a quantum field theory
 is rather loaded. Hence first we shall clarify some terms and conventions that
 will be used in the rest of this article. In the following we consider
 models where only scalars can get a \vev as required by Lorentz invariance.

In principle, the effective potential is a
 real-valued\footnote{As noted in \Sec~\ref{sec:convexity},
 the loop expansion may lead to complex values for the one-loop effective
 potential. See \cite{Fujimoto:1982tc, Weinberg:1987} for detailed
 discussions.} functional over all the quantum fields of the model.
 However, under the assumption that the vacuum is homogeneous and
 isotropic, for the purposes of determining the vacua of the model, the
 effective potential can be treated as a function of sets of (dimensionful)
 numbers, which we shall refer to as \textit{field configurations}. Each field
 configuration is a set of variables which correspond to the classical
 expectation values for the spin-zero fields which are constant with respect to
 the spatial co-ordinates.

The example of the SM is relatively simple: the field configuration is simply a
 set of two complex numbers, which are the values of the neutral and charged
 scalar fields assuming that each is constant over all space. These four real
 degrees of freedom can be reduced to a single degree of freedom by employing
 global $SU(2)_{L}$ and phase rotations, leaving an effective potential that is
 effectively a function of a single variable.

Henceforth we shall assume that each complex field is treated as a pair of real
 degrees of freedom, and note that this may obscure continuous sets of
 physically equivalent degrees of freedom which are manifestly related by phase
 rotations when expressed with complex fields.

Also we shall refer to the local minima of the effective potential as its
 \textit{vacua}, and label the global minimum as the \textit{true} vacuum, while
 all the others are \textit{false} vacua. A potential may have multiple true
 vacua, either as a continuous set of minima related by gauge transformations as
 in the SM for example, or a set of disjoint, physically inequivalent minima,
 each of which may of course be a continuous set of physically equivalent minima
 themselves. In cases where there is a continuous set of physically equivalent
 minima, we assume that a single exemplar is taken from the set for the purposes
 of comparison of physically inequivalent minima.

Furthermore, the term \textit{vacuum expectation value} can be used in many
 confusing ways. In this work, \vevs will only refer to the sets of constant
 values which the scalar fields have at the field configurations which minimize
 the effective potential. Hence we do {\em not} consider the effective potential
 to be a ``function of the \vevs'', rather a function of a set of numbers that
 we call a field configuration.

\subsection{The tree-level potential and tadpole equations}

If one merely considers the SM at tree level, 
 minimizing the potential is straightforward. A global
 $SU(2)_{L}$ rotation can bring the part of the potential due to the scalar
 doublet into the form
\begin{equation}
 V = \frac{\lambda}{4} | \phi |^{4} + \frac{{\mu}^{2}}{2} | \phi |^{2}
\end{equation}
 where $\phi$ is the neutral component of the $SU(2)_{L}$ doublet. After a
 little differentiation and algebra, one finds that if
 $\lambda> 0, {\mu}^{2} < 0$, then the potential is minimized for
 $| \phi | = v = \sqrt{-{\mu}^{2} / \lambda}$.

However, for a set of $N$ real scalar degrees of freedom ${\phi}_{i}$ (writing
 complex scalar fields as two separate real scalars), the scalar part of
 the tree-level potential of a renormalizable quantum field theory in four
 space-time dimensions is of the form
\begin{equation}
\label{eq:tree_level_potential}
 V^{\text{tree}}
 = {\lambda}_{ijkl} {\phi}_{i} {\phi}_{j} {\phi}_{k} {\phi}_{l}
   + A_{ijk} {\phi}_{i} {\phi}_{j} {\phi}_{k}
   + {\mu}^{2}_{ij} {\phi}_{i} {\phi}_{j}
   + \text{irrelevant constant term}
\end{equation}
 which, when differentiated with respect to the $N$ independent ${\phi}_{i}$,
 yields $N$ polynomial equations of up to degree three. We have assumed that any
 terms linear in the fields have been removed by constant shifts of the fields.

Although we assume renormalized potentials here for simplicity, the methods
 used by \vcs are equally applicable to non-renormalizable potentials, as long
 as $V^{\text{tree}}$ is expressed as a finite-degree polynomial. The value of
 loop corrections to a non-renormalizable potential may be debatable, but \vcs
 can be restricted to using just the tree-level potential.

While closed-form solutions for cubic polynomials in one variable exist, solving
 a coupled system in general requires very involved algorithms, such as using
 Gr{\"{o}}bner bases to decompose the system \cite{Maniatis:2006jd,Gray:2008zs},
 or homotopy continuation to trace known solutions of simple systems as
 they are deformed to the complicated target system of tadpole equations.

\subsubsection{The homotopy continuation method}

The homotopy continuation method \cite{sommesenumerical, li2003numerical}
 has found use in several areas of physics
 \cite{Mehta:2009zv, Mehta:2011xs, Mehta:2012qr}, in particular to find string
 theory vacua \cite{Mehta:2011wj, Mehta:2012wk} and extrema of extended Higgs
 sectors \cite{Maniatis:2012ex}, where the authors investigated a system of two
 Higgs doublets with up to five singlet scalars in a general tree-level
 potential, and \cite{CamargoMolina:2012hv}, where systems of up to ten fields
 were allowed to have non-zero \vevs. In contrast, the Gr{\"{o}}bner basis
 method is deemed prohibitively computationally expensive for systems involving
 more than a few degrees of freedom \cite{Maniatis:2006jd}.

The numerical polyhedral homotopy continuation method is a powerful way to
 find all the roots of a system of polynomial equations quickly
 \cite{lee2008hom4ps}. Essentially it works by continuously deforming a simple
 system of polynomial equations with known roots, with as many roots as the
 classical B{\'{e}}zout bound of the system that is to be solved (\ie~the
 maximum number of roots it could have). The simple system with known roots is
 continuously deformed into the target system, with the position of the roots
 updated with each step. While the method is described in detail in
 \cite{sommesenumerical, li2003numerical}, a light introduction can be found for
 example in \cite{Maniatis:2012ex}.

\subsection{The one-loop potential}

\newcommand{\MM}[0]{\ensuremath{{\bar{M}}^{2}_{n}( \Phi )}\xspace}

The general form of the renormalized one-loop effective potential
 \cite{Sher:1988mj, Martin:2002} is
\begin{equation}
\label{eq:potential_in_parts}
 V^{\text{1-loop}}
 = V^{\text{tree}} + V^{\text{counter}} + V^{\text{mass}}
\end{equation}
 where $V^{\text{tree}}$ is as above and $V^{\text{counter}}$ has the same form
 as $V^{\text{tree}}$, \ie~a polynomial of the same degree in the same fields,
 but the coefficients are instead the renormalization-dependent finite
 parts of the appropriate counterterms \cite{Sher:1988mj}.
The term $V^{\text{mass}}$ has the form, for a given field configuration $\Phi$,
\begin{equation}
\label{eq:potential_loop_corrections}
 V^{\text{mass}} = \frac{1}{64 {\pi}^{2}}
 \sum_{n} (-1)^{(2 s_{n})} ( 2 s_{n} + 1 ) ( {\MM} )^{2}
 \left[ \log( \MM / Q^{2} ) - c_{n} \right]
\end{equation}
 where the sum runs over all real scalar, Weyl fermion, and vector degrees of
 freedom, with $s_{n}$ being the spin of the degree of freedom. Complex scalars
 and Dirac fermions are accounted for as mass-degenerate pairs of real scalars
 and Weyl fermions respectively.

For scalar degrees of freedom, the \MM are the eigenvalues of the second
 functional derivative of $V^{\text{tree}}$, \ie~the eigenvalues of
 $({\bar{M}}^{2}_{s=0})_{ij}
 = ( {\lambda}_{ijkl} + {\lambda}_{ikjl} + {\lambda}_{iklj} + ... )
   {\phi}_{k} {\phi}_{l}
   + ( A_{ijk} + A_{ikj} + A_{kij} + ... ) {\phi}_{k}
   + {\mu}^{2}_{ij} + {\mu}^{2}_{ji}$. Thus these \MM are the eigenvalues of the
 tree-level scalar ``mass-squared matrix'' that would be read off the Lagrangian
 with the scalars written as fluctuations around the field configuration. Unless
 the field configuration corresponds to a minimum of the effective potential,
 these do not correspond to physical masses in any way, of course.

Likewise, the \MM for fermionic and vector degrees of freedom are the
 eigenvalues of the respective ``mass-squared matrices'' where the scalar fields
 are taken to have constant values given by the field configuration. (The
 fermion mass-squared matrix is given by the mass matrix multiplied by its
 Hermitian conjugate.)

The terms $c_{n}$ depend on the regularization scheme. In the \msbar
 scheme, $c_{n}$ is $3/2$ for scalars and Weyl fermions, but $5/6$ for vectors,
 while in the \drbarp scheme \cite{Jack:1994rk, Martin:2002}, more suitable for
 supersymmetric models, $c_{n}$ is $3/2$ for all degrees of freedom. Since this
 is a finite-order truncation of the expression, the renormalization scale $Q$
 also appears explcitly in the logarithm, as well as implicitly in the scale
 dependence of the renormalized Lagrangian parameters.

Much of the literature on one-loop potentials (including \cite{Martin:2002})
 assumes a renormalization scheme where $V^{\text{counter}}$ is zero; however,
 such a scheme is often inconvenient for other purposes, such as ensuring
 tadpole equations have a given solution at the one-loop level,
  see \eg the appendix of \cite{Hirsch:2012kv}.
 (\sarah automatically generalizes this approach to extended SUSY models as
 explained in \cite{Staub:2010ty, OLeary:2011yq}.)
Finally, we also note that models can be constructed where spontaneous symmetry
 breaking does not happen at tree level, but does exist when one takes loop
 corrections into account \cite{Coleman:1973jx, Fujimoto:1982tc}.

\subsubsection{Scale dependence}
\label{sec:scale_dependence}

As noted, the one-loop effective potential depends on the renormalization scale.
 Ideally one would use the ``renormalization-group improved'' expression for
 the potential \cite{Sher:1988mj} as this is invariant under changes of scale;
 however, this is often totally impractical except for potentials with only a
 handful of parameters and a single scalar field.

If one must use a scale-dependent expression for the potential, as is often the
 case, the renormalization scale should be chosen carefully: if one chooses a
 scale too high or too low, one may find that with a finite-loop-order (and thus
 scale-dependent) effective potential, there is no spontaneous breaking of any
 symmetry, or even that the potential is not bounded from below
 \cite{Gamberini:1989jw}! This is often simply due to the fact that higher
 orders become more important in such a case, especially when corrections from
 the next order would introduce new, large couplings, such as often happens when
 going from tree level to one loop. It can also be that the scale is so large or
 small that the loop expansion is no longer a reliable expansion. We also note
 that rather undermines arguments that radiative effects do not change
 tree-level conclusions on the absolute stability of vacua such as in
 \cite{PhysRevD.48.4352} (where the argument also fails to take into account
 that there may not even be a scale at which the renormalization condition used
 can be satisfied).

Indeed, it is crucial that the scale is chosen so that the loop expansion is
 valid. Explicitly, large logarithms should not spoil the perturbativity of the
 expansion in couplings. Loops with a particle $n$ typically come with a factor
 of $\ln( \MM / Q^{2} )$ along with the factor of
 ${\alpha}_{n} = [$relevant coupling$]^{2} / ( 4 \pi )$, and thus
 ${\alpha}_{n} \ln( \MM / Q^{2} )$ should remain sufficiently smaller than one
 such that the expansion can be trusted \cite{Sher:1988mj}. A rough first
 estimate then of the region of validity, assuming that the dimensionful
 Lagrangian parameters are all of the order of the renormalization scale to some
 power, is where $\ln( v^{2} / Q^{2} ) / ( 4 \pi ) \leq 1 / 2$, say, for a field
 configuration with vector length $v$, so where the \vevs are within a factor of
 $e^{\pi} \simeq 20$ of the renormalization scale.

Furthermore, it is in general not valid to compare the potential for different
 field configurations using a different scale for each configuration, if one is
 using a scale-dependent effective potential. The reason is that there is an
 important contribution to the potential that is field-independent yet still
 depends on the scale\footnote{For example, if one is comparing two field
 configurations of the MSSM potential where the squark fields happen to have
 zero values, the mass of the gluino is independent of the non-zero fields and
 yet provides a scale-dependent contribution to the effective potential.}
\cite{Ferreira:2000hg, Einhorn:2007rv}. Of course, if one knows the full scale
 dependence of all the terms of the Lagrangian regardless of whether they lead
 to field-dependent contributions to the effective potential, then one can
 correctly evaluate different field configurations at different scales.

\subsection{Gauge dependence}
\label{sec:gauge_dependence}

The one-loop potential is explicitly gauge-dependent
 \cite{PhysRevD.9.2904, Nielsen:1975fs}. However, as shown in
 \cite{Nielsen:1975fs, Fukuda:1975di}, the values it takes at its extrema are
 independent of the gauge chosen, except for spurious extrema of poorly-chosen
 gauges. The popular $R_{\xi}$ gauges are well-behaved and do not have
 fictitious gauge-dependent extrema for reasonable choices of $\xi$
\cite{Fukuda:1975di}.

It is also possible to formulate the effective potential in terms of
 gauge-invariant composite fields \cite{Buchmuller:1994vy}, though this may not
 always be practical. One can also verify the gauge-independence of extrema
 using more complicated gauges and applying BRST invariance 
\cite{Kastening:1993zn}.

\subsection{Convexity}
\label{sec:convexity}

As shown in \cite{Fujimoto:1982tc}, the effective potential, which can be
 thought of as the quantum analogue of the classical potential energy for
 constant fields, is real for all values of the fields. However, the loop
 expansion leads to complex values in regions where the classical potential is
 non-convex. While one can take the convex hull of the truncated expansion of
 the potential when evaluating the potential for configurations of fields in the
 convex region, it is not particularly helpful for the purpose of computing
 tunneling transition times. Fortunately, the one-loop truncation of the
 effective potential as a function of constant values for the fields can
 consistently be interpreted as a complex number with real part giving the
 expectation value of the potential energy density for the given field
 configuration and imaginary part proportional to the decay rate per unit volume
 of this configuration \cite{Weinberg:1987}.

\subsection{Comparing two vacua}

If there are two or more physically inequivalent minima of a potential, then it
 is vitally important to know if the phenomenologically desired minimum is the
 global minimum, or, if not, how long the expected tunneling time to the true
 vacuum is.

Given the issues raised above in sections \ref{sec:scale_dependence} and
 \ref{sec:gauge_dependence}, it is safe to use a scale- and gauge-dependent
 one-loop effective potential to compare two inequivalent minima provided that
 the scale is held fixed and that the two minima are within the region of
 validity determined by the renormalization scale. Of course, an explicitly
 gauge- and scale-independent expression for the effective potential would
 obviously be unburdened by such concerns, but unfortunately it is rare to be
 able to formulate such an expression.

\section{Tunneling from false vacua to true vacua}
\label{sec:tunneling}

The usual expression for the decay rate $\Gamma$ per unit volume for a false
 vacuum is given in \cite{Coleman:1977py, Callan:1977pt} as
\begin{equation}
 \Gamma / \text{vol.} = A e^{( -B / \hbar)} ( 1 + \mathcal{O}( \hbar ) )
\label{eq:tunneling_time}
\end{equation}
 where $A$ is a factor which depends on eigenvalues of a functional determinant
 and $B$ is the bounce action. The $A$ factor is typically estimated on
 dimensional grounds as it is very complicated to calculate and, because of the
 exponentiation of $B$, is far less important than getting the bounce action as
 accurate as possible. If $A$ is taken to be
 $\mathcal{O}( ( 100 - 1000 \gev )^{4} )$, then for $\Gamma / \text{vol.}$ to be
 roughly the age of the known Universe to the fourth power, $B$ must be around
 $400 \hbar$, and a per-cent variation in $B$ leads to a factor of $e$ variation
 in the tunneling time.

Given a path through the field configuration space from one vacuum to another
 with a lower value, which for convenience we shall label as the false vacuum
 and true vacuum respectively, one can solve the equations of motion for a
 bubble of true vacuum of critical size in an infinite volume of false vacuum
 \cite{Coleman:1977py, Wainwright:2011kj}. This allows one to calculate the
 bounce action and thus the major part of the tunneling time.

Unfortunately, this means that to calculate the tunneling time from a false
 vacuum to a true vacuum, one needs to evaluate the potential along a continuous
 path through the field configuration space, and even though the extrema of the
 potential are gauge-invariant as noted above, the paths between them are not.
 However, it has been proved that at zero temperature, the gauge dependence at
 one-loop order cancels out \cite{Metaxas:1995ab}.

 At finite temperature, the situation is not so clear, though the Landau gauge
 may be most appropriate \cite{PhysRevD.11.905}.
 While some studies have shown that for ``reasonable'' choices of gauge, the
 differences in finite-temperature tunneling times are small
 \cite{Garny:2012cg}, it is still possible to choose poor gauges that can even
 obscure the possibility of tunneling \cite{Wainwright:2012zn}.

\section{\vcs: objectives, outline, features and limitations}
\label{sec:features}

\subsection{Objectives}

The \vcs program is intended as a tool to quickly evaluate whether a parameter
 point with a given set of \vevs, referred to henceforth as the
 \textit{input vacuum}, has, to one-loop order\footnote{As mentioned in
 \Sec~\ref{sec:scale_dependence}, a renormalization-group-improved effective
 potential would be better than the one-loop effective potential, but at the
 moment it seems infeasible to implement.}, any vacua with lower potential
 energy than the input vacuum, and, optionally, to estimate the tunneling time
 from the input vacuum to the true vacuum if so.

A typical use envisaged is a parameter scan for a single model. Some effort
 needs to be put into creating the model file in the first place, though this is
 straightforward if using \sarah as described in section \ref{sec:sarah_input};
 once the model file has been created, parameter points given in the form
 of \slha files should be evaluated within a matter of seconds, depending on how
 complicated the model is, what simplifications have been made, and how
 accurately the tunneling time should be calculated if necessary.

Given a model (through a model file) and a parameter point (through an \slha
 file), \vcs determines the global minimum of the one-loop effective potential,
 and a verdict on whether the input minimum is absolutely stable, by it being
 the global minimum, or metastable. The user provides a threshold for which
 the metastability is rated as long-lived or short-lived. Whether the tunneling
 time or just an upper bound is calculated depends on whether the upper bound is
 above or below the threshold, or may be forced by certain options (see
 \Sec~\ref{sec:options}).

\subsection{Outline}
\label{sec:vcs_outline}

\begin{figure}
\begin{tikzpicture}[scale=1.0, transform shape]

\draw[ fill = blue!10 ] ( 0.0, 0.0 ) rectangle ( 8.0, 20.0 );
\node[ anchor = west ] (vcsExe) at ( 0.2, 19.6 ) {
 {\tt Vevacious.exe} (from {\tt Vevacious.cpp}) };

\draw[ fill = blue!05 ] ( 0.2, 0.2 ) rectangle ( 7.8, 19.15 );
\node[ anchor = west ] (vcsRunner) at ( 0.4, 18.9 ) {
 {\tt Vevacious::VevaciousRunner} };

\node[ fill = red!30, draw, rectangle ] (modelFile) at ( 10.5, 19.5 ) {
 {\tt MyModel.vin} };

\node[ fill = red!30, draw, rectangle ] (initStuff)
 at ( 11.5, 18.1 ) {
 \makecell[l]{{\tt VevaciousInitialization.xml}\\
 and/or command-line arguments} };

\node[ fill = red!30, draw, rectangle ] (slhaFile) at ( 0.5, 14.8 ) {
 {\tt MyParameters.slha.out} };

\node[ fill = blue!01, draw, rectangle ] (vcsRunnerCtor)
 at ( 4.0, 18.3 ) {
 constructor };
\path[->] [red] (modelFile) edge [ out = 180, in = 0 ] (vcsRunnerCtor);
\path[->] [red] (initStuff) edge [ out = 180, in = 0 ] (vcsRunnerCtor);
\node[ fill = blue!01, draw, rectangle ] (vcsRunnerSetup)
 at ( 4.0, 17.7 ) {
 setters };
\path[->] [red] (initStuff) edge [ out = 180, in = 0 ] (vcsRunnerSetup);
\node[ fill = blue!01, draw, rectangle ]
 (findExtrema) at ( 4.0, 16.4 ) {
 \makecell[l]{{\tt findTreeLevelExtrema(} \\
 {\tt std::string const\& slhaFilename )}} };
\path[->] [red] (slhaFile) edge [ out = 0, in = -90 ] (findExtrema);

\draw[ fill = green!10 ] ( 8.2, 12.5 ) rectangle ( 15.7, 17.0 );
\node[ fill = green!05, draw, rectangle ] (cdHomps)
 at ( 12.0, 16.6 ) {
 change directory to location of \homps };
\node[ fill = green!05, draw, rectangle ] (runHomps)
 at ( 12.0, 15.8 ) {
 write input file and run \homps };
\node[ fill = green!05, draw, rectangle ] (parseHomps)
 at ( 12.0, 14.3 ) {
 \makecell[l]{parse output of \homps, \\
 discard invalid and duplicate solutions} };
\node[ fill = green!05, draw, rectangle ] (cdOriginal)
 at ( 12.0, 12.9 ) {
 return to original directory };
\path[->] [green!50!black] (findExtrema) edge [ out = 0, in = 180 ] (cdHomps);
\path[->] [green!50!black] (cdHomps) edge [ out = -90, in = 90 ] (runHomps);
\path[->] [green!50!black] (runHomps) edge [ out = -90, in = 90 ] (parseHomps);
\path[->] [green!50!black] (parseHomps) edge [ out = -90, in = 90 ] (cdOriginal);

\node[ fill = red!30, draw, rectangle ] (extremaPy)
 at ( 13.0, 11.5 ) {
 {\tt VevaciousTreeLevelExtrema.py} };
\path[->] [red] (parseHomps) edge [ out = -20, in = 20 ] (extremaPy);

\node[ fill = blue!01, draw, rectangle ]
 (preparePython) at ( 4.0, 12.5 ) {
 \makecell[l]{{\tt prepareParameterDependentPython(} \\
 {\tt std::string const\& slhaFilename )}} };
\path[->] [red] (slhaFile) edge [ out = 0, in = 90 ] (preparePython);

\node[ fill = red!30, draw, rectangle ] (potentialPy)
 at ( 9.0, 10.6 ) {
 {\tt VevaciousParameterDependent.py} };
\path[->] [red] (preparePython) edge [ out = -90, in = 180 ] (potentialPy);

\node[ fill = blue!01, draw, rectangle ]
 (runPython) at ( 2.0, 11.1 ) {
 {\tt runPython()} };

\draw[ fill = green!10 ] ( 0.45, 3.0 ) rectangle ( 15.8, 9.9 );
\node[ fill = green!05, draw, rectangle, anchor = west ] (runVcsPy)
 at ( 2.0, 9.5 ) {
 run {\tt Vevacious.py}, ensuring it exists: };
\path[->] [green!50!black] (runPython) edge [ out = -90, in = 180 ] (runVcsPy);
\draw[ fill = yellow!20 ] ( 0.7, 3.1 ) rectangle ( 15.6, 9.1 );
\node[ anchor = west ] (vcsExe) at ( 0.8, 8.8 ) {
 {\tt Vevacious.py} };
\node[ fill = yellow!05, draw, rectangle, anchor = west ] (createPym)
 at ( 1.0, 8.3 ) {
 create \pyminuit object for effective potential function };
\path[->] [red] (potentialPy) edge [ out = -90, in = 5 ] (createPym);
\node[ fill = yellow!05, draw, rectangle, anchor = west ] (forExtrema)
 at ( 1.0, 7.6 ) {
 run \pyminuit for each tree-level extremum };
\path[->] [red] (extremaPy) edge [ out = -90, in = 0 ] (forExtrema);
\node[ fill = yellow!05, draw, rectangle, anchor = west ] (checkSaddle)
 at ( 1.5, 6.8 ) {
 (if \pyminuit stops at a saddle point, nudge it off ${}^{\dagger}$) };
\node[ fill = yellow!05, draw, rectangle, anchor = west ] (sortMinima)
 at ( 1.0, 5.9 ) {
 sort minima and compare to input minimum };
\node[ fill = yellow!05, draw, rectangle, anchor = west ] (quickCt)
 at ( 1.0, 5.0 ) {
 if required, get ``direct path'' tunneling time upper bound with \ct
 ${}^{\dagger}$ };
\node[ fill = yellow!05, draw, rectangle, anchor = west ] (fullCt)
 at ( 1.0, 4.2 ) {
 if required, get ``deformed path'' tunneling time with \ct
 ${}^{\dagger}$ };
\node[ fill = yellow!05, draw, rectangle, anchor = west ] (writeXml)
 at ( 1.0, 3.5 ) {
 write results };

\node[ fill = red!30, draw, rectangle ] (resultsXml)
 at ( 11.0, 2.5 ) {
 {\tt MyResults.vout} };
\path[->] [red] (writeXml) edge [ out = 0, in = 180 ] (resultsXml);

\node[ fill = blue!01, draw, rectangle ]
 (appendSlha) at ( 4.0, 1.5 ) {
 \makecell[l]{{\tt appendResultsToSlha(} \\
 {\tt std::string const\& slhaFilename )}} };
\path [red][<-] (slhaFile) edge [ out = -100, in = 90 ] ( -0.2, 3.0 );
\path [red][->] ( -0.2, 3.0 ) edge [ out = -90, in = 180 ] (appendSlha);
\path[->] [red] (resultsXml) edge [ out = -90, in = 0 ] (appendSlha);

\end{tikzpicture}
\caption{\vcs flow diagram.
 The member functions
 of \texttt{VevaciousRunner} are shown from top to bottom in the order in which
 they are called by \texttt{Vevacious.exe}, as can be seen by looking into the
 \texttt{Vevacious.cpp} source file. All steps labeled with ${}^{\dagger}$ are
 elaborated in the text.
 (Most file names can be changed, see \Sec~\ref{sec:options}).}
\label{fig:flow_diagram}
\end{figure}
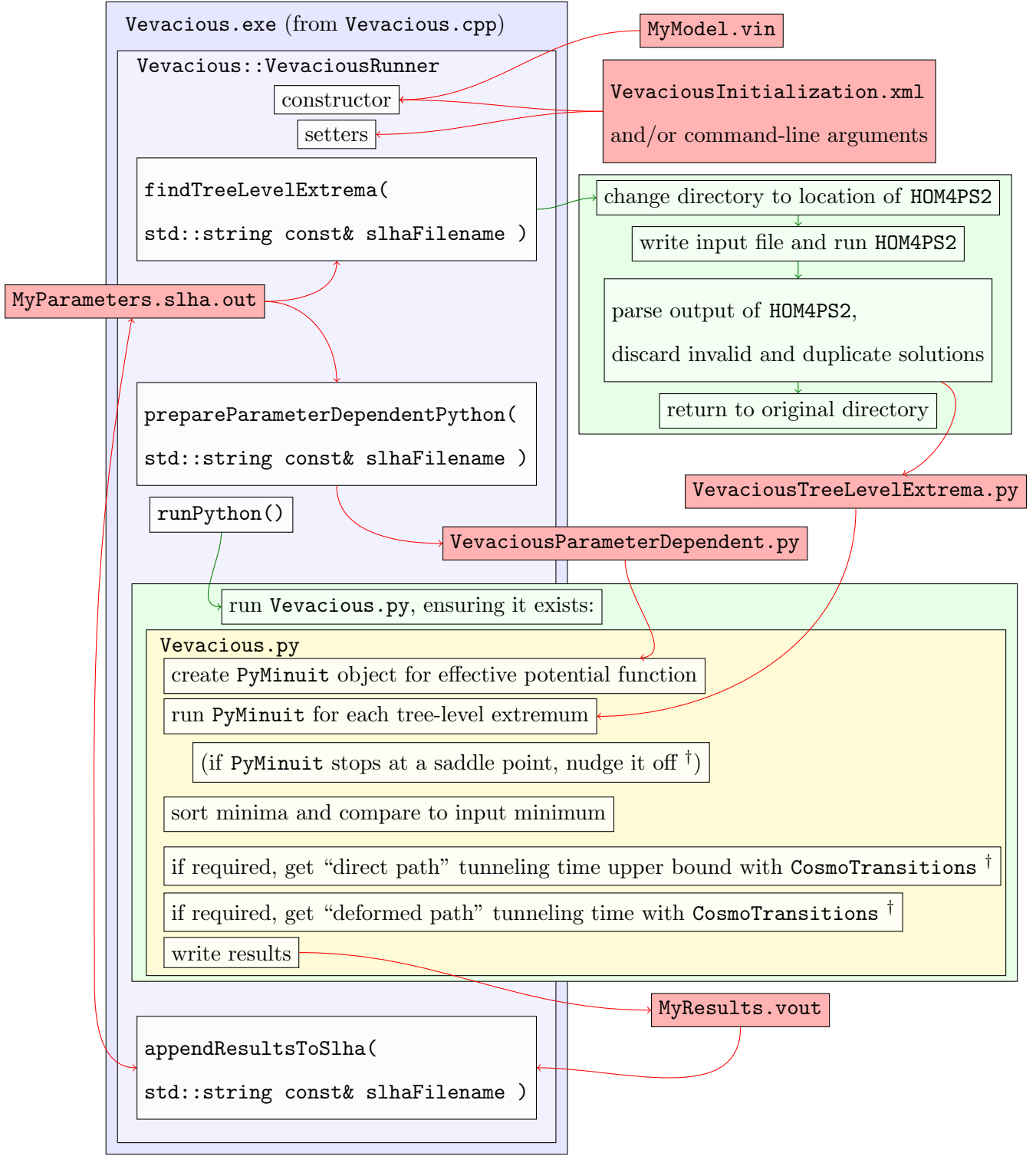

Here we present the steps taken by \vcs, which are schematically shown in
 \fig~\ref{fig:flow_diagram}.

\begin{itemize}

 \item[(1)] An input file in the \slha format \cite{Skands:2003cj} is read
 in to obtain the Lagrangian parameters defining the parameter point, required
 to evaluate the potential. We emphasize that even though the {\it SUSY} Les
 Houches Accord is used as the format, the model itself does not need to be
 supersymmetric, as long as the \slha file contains appropriate \blocks.

 \item[(2)] {\em All} the extrema of the tree-level potential are found
 using the homotopy continuation method \cite{sommesenumerical} to solve the
 tree-level tadpole equations. The publicly available program \homps
 \cite{lee2008hom4ps} is used for this.
 
 \item[(3)] The tree-level extrema are used as starting points for
 gradient-based minimization of the one-loop effective potential. The \minuit
 algorithms \cite{James:1975dr} are used here through the \py wrapper \pyminuit
 \cite{pyminuit}. The points where \pyminuit stops are checked to see if they
 are really minima\footnote{\pyminuit stops when it has found a sufficiently
 flat region without checking whether it is at a minimum.}. Any saddle point is
 then split into two further points, displaced from the original in the
 directions of steepest descent by amounts given by the \xmle{saddle\_nudges}
 arguments (see \Sec~\ref{sec:options}), which are then also used as starting
 points for \pyminuit. By default, \pyminuit is restricted to a hypercube of
 field configurations where each field is only allowed to have a magnitude less
 than or equal to one hundred times the renormalization scale of the \slha file.
 This is rather excessive by the reasoning of \Sec~\ref{sec:scale_dependence},
 which would lead one to take at most maybe ten or twenty times the scale as an
 upper limit; however, it was considered better to allow the user to decide
 whether the results of \vcs are within a trustworthy region.
 
 \item[(4)] The minima are sorted, and, if necessary, the tunneling time
 from the input vacuum to the true vacuum is calculated. The $A$ factor of
 \eq~(\ref{eq:tunneling_time}) is taken to be equal to the fourth power of the
 renormalization scale, as this is expected to be the typical scale of the
 potential and thus the expected scale of the solitonic solutions, and the
 bounce action is calculated with the code \cosmotransitions
 \cite{Wainwright:2011kj}. To save time, first \cosmotransitions is called to
 calculate the bounce action with a bubble profile given by a straight line in
 field configuration space from the false vacuum to the true vacuum to get an
 upper bound on the tunneling time. If this upper bound is below the user-given
 threshold \xmle{direct\_time} (see \Sec~\ref{sec:options}), then no refinement
 is pursued. If, however, the upper bound is above the threshold, the bounce
 action is calculated again allowing \cosmotransitions to deform the path in
 field configuration space to find the minimal surface tension for the bubble.
 If one wishes to calculate the tunneling time with a different $A$ factor, one
 can edit a line of \py code as described in \Sec~\ref{subsec:features}.

 \item[(5)] The results are printed in a results file and also appended to
 the \slha input file.

\end{itemize}

\subsection{Features}
\label{subsec:features}

\paragraph{Finds all tree-level extrema.}
The homotopy continuation method is guaranteed to find all the solutions of the
 system of tadpole equations (to the limitations of the finite precision of the
 machine following the algorithm) \cite{sommesenumerical}. One does not have to
 worry that there may be solutions just beyond the range of a scan looking for
 the solutions.

\paragraph{Rolls to one-loop minima.}
 \vcs rolls from the tree-level extrema to the minima of the one-loop effective
 potential before comparing them, because in general the \vevs get shifted. In
 addition, extrema that change their nature with radiative corrections, such as
 the field configuration with zero values for all the fields in the
 Coleman--Weinberg model of radiative spontaneous symmetry breaking
 \cite{Coleman:1973jx}, which is a minimum of the tree-level potential but a
 local maximum of the one-loop effective potential, are found.

\paragraph{Calculates tunneling times or upper bounds on them.}
A parameter point in a model is not necessarily ruled out on the basis that the
 desired minimum of the potential is not the global minimum, since a metastable
 configuration with a lifetime of roughly the observed age of the Universe or
 longer is compatible with the single data point that we have. \vcs creates
 \cosmotransitions objects with its effective potential function to evaluate the
 bounce action and thus the tunneling time from a false vacuum to the true
 vacuum of a parameter point.

\paragraph{Fast.}
An important aspect of \vcs is that it is fast enough to be used as a check in
 a parameter scan of a model. For example, on a laptop with a $2.4$ GHz
 processor, a typical parameter point for the MSSM allowing
 six real non-zero \vevs (two Higgs, two stau, two stop) can report within $3.2$
 seconds that no deeper vacuum than the input vacuum was found, or, for a
 different parameter point, can report an upper bound on the tunneling time
 within 18 seconds. However, borderline cases which require a full calculation
 of the minimal bounce action can take up to 500 seconds. Reducing the number of
 degrees of freedom to four (fixing the stop values at zero) reduces the
 calculation times to $0.6$, $2.3$ and $27$ seconds respectively.

\paragraph{Flexible.}
\vcs has been written in a way that should allow useful customizations with
 small changes to the main \py code. For example, one can change a single line
 (line 36) in \texttt{Vevacious.py}
 so that the tree-level potential is used for the analysis rather than
 the one-loop effective potential:
\begin{lstlisting}
effectivePotentialFunction = VPD.LoopCorrectedPotential
\end{lstlisting}
 can be changed to
\begin{lstlisting}
effectivePotentialFunction = VPD.TreeLevelPotential
\end{lstlisting}
 and no further changes are necessary. \texttt{Vevacious.exe} does not overwrite
 \texttt{Vevacious.py}, so any changes to the \py code will be kept. This was
 chosen as the best compromise to allow non-trivial changes without forcing the
 user to go very deep into the code, though it does rely on the user learning
 some \py to be able to do so. Another customization that one may wish to make
 could be to change the $A$ factor for the calculation of the tunneling time
 (and hence the thresholds for the bounce action calculations). This would be
 done by editing line 433 of \texttt{Vevacious.py} to fix the fourth root of $A$
 from the renormalization scale:
\begin{lstlisting}
fourthRootOfSolitonicFactorA = VPD.energyScaleFourth
\end{lstlisting}
 can be changed to
\begin{lstlisting}
fourthRootOfSolitonicFactorA = 246.0
\end{lstlisting}
 to change the $A$ factor to $(246$ GeV$)^{4}$ for example, if one feels that
 the electroweak scale is a more appropriate choice.

\subsection{Limitations}
\label{subsec:limitations}

\paragraph{Garbage in, garbage out.}
\vcs performs very few sanity checks, so rarely protects the user from their
 own mistakes. For instance, \vcs does not check if the potential is bounded
 from below. However, there are some checks, such as those which result in the
 warning that the given input minimum was actually rather far from the nearest
 minimum found by \vcs (though \vcs carries on regardless after issuing the
 warning).  Another important sanity check that is {\em not} performed is to
 check that the \slha \blocks required by the model file are actually present in
 the given \slha input file. The user is fully responsible for providing a valid
 \slha file to match the model. As model files are expected to be produced
 automatically by software such as \sarah, it is expected that the \slha files
 for the model will also be prepared consistently with the expected \blocks.
 Unfortunately it is quite easy to miss this point when using the example model
 files provided by default with \vcs: if one does use these files, one must use
 the correct model file for the input \slha files, as described in
 \Sec~\ref{sec:SUSY_examples}.

\paragraph{May be excessively optimistic about the region of validity.}
By default, \vcs allows \vevs to have values up to a hundred times the
 renormalization scale, and it is up to the user to decide whether any given
 set of results is meaningful and within the region of validity of the one-loop
 effective potential used. However, it is straightforward to change the allowed
 region to a smaller multiple of the renormalization scale by editing line 85 of
 the default {\tt Vevacious.py} from
\begin{lstlisting}
    minuitObject.limits[ vevVariable ] = ( -100.0, 100.0 )
\end{lstlisting}
 to, for example,
\begin{lstlisting}
    minuitObject.limits[ vevVariable ] = ( -20.0, 20.0 )
\end{lstlisting}
 to limit \vevs to be no larger than twenty times the renormalization scale. One
 could also insert more complicated \py code here, but one should be aware that
 the \pym object deals with a potential where the field values are in units of
 the renormalization scale.

\paragraph{Not guaranteed to find minima induced purely by radiative effects.}
While \vcs does find all the extrema of the tree-level effective potential,
 there is no guarantee that these correspond to all the minima of the effective
 potential at the one-loop level. The strategy adopted by \vcs will find all the
 minima of the one-loop effective potential that are in some sense ``downhill''
 from tree-level extrema, but any minima that develop which would require
 ``going uphill'' from every tree-level extremum will not be found. Such
 potentials are not impossible: if the quadratic coefficient in the
 Coleman-Weinberg potential \cite{Coleman:1973jx} is small enough while still
 positive, the single tree-level minimum can remain a minimum at the one-loop
 level while deeper minima induced by radiative corrections still appear.
 However, if the tree-level minimum is sufficiently shallow then the finite
 numerical derivatives used by \vcs may be enough to push it over the small
 ``hills'' into the one-loop minima.

\paragraph{Extreme slow-down with too many degrees of freedom.}
Like many codes, the amount of time \vcs needs to produce results increases
 worse than linearly with the number of degrees of freedom. A proper
 quantification of exactly how \vcs scales with degrees of freedom remains on
 the to-do list, but as a guide, some typical running times
 (again on a $2.4$ GHz core) for the \homps part
 are: 3 degrees of freedom: $0.03$ seconds; 5: $0.28$ seconds; 7: $5.1$ seconds;
 10: 20 minutes; 15: 10 days. The \pyminuit part depends on the number of
 solutions found by \homps, but in general takes several seconds. The \ct part
 is strongly dependent on the details of a particular potential, and how rapidly
 the path deformations converge; models with the same degrees of freedom can
 vary wildly from seconds to hours to be computed. For this reason, \vcs only
 calls the full calculation of \ct if the quick estimate of the upper bound on
 the tunneling time is over the user-given threshold, and also gives the user
 the option to never use the full calculation, rather only the quick upper-bound
 calculation, which in general takes only a few seconds at most.

\paragraph{Homotopy continuation method requires discrete extrema.}
The homotopy continuation method relies on tracking the paths of a discrete
 number of simple solutions to a discrete number of target solutions. There is
 no guarantee that a system with a continuous set of degenerate solutions will
 be solved by \homps, and unfortunately \vcs can not check that the system has
 redundant degrees of freedom such as those corresponding to a gauge
 transformation. Hence the user must choose the degrees of freedom of the model
 appropriately.

\paragraph{Homotopy continuation path tracking resolution.}
The homotopy continuation method guarantees that there is a path from each
 solution of the simple system to its target solution, however, there is a
 danger that a finite-precision path-tracking algorithm will accidentally
 ``jump'' from the path it should be following onto a very close other path to a
 different solution, possibly leading to one or more solutions remaining unfound
 (though not necessarily, since several simple solutions may map to the same
 (degenerate) target solution).

\paragraph{Tunneling path resolution.}
Calculating the tunneling time requires finding a continuous path in field
 configuration space from the false to the true vacuum. However, this must be
 discretized to a finite number of points on a finite machine, and may even
 lead to a very small barrier between the vacua disappearing entirely. However,
 in such cases the tunneling time should be very short indeed, so \vcs notes
 this and takes a fixed very small tunneling time as the result.

\section{Subtleties: renormalization schemes and allowed degrees of freedom}
\label{sec:subtleties}

Currently, \vcs performs very few sanity checks. In particular, it remains
 blissfully ignorant of any physical meanings the user intends for the values of
 the Lagrangian parameters which are given. Thus is it entirely up to the user
 to ensure that these values correspond correctly to the intended
 renormalization conditions. However, \vcs \textit{does} assume
 some form of dimensional regularization (\eg switching between \drbarp and
 \msbar depends on the value for $c_{n}$ given for the vector mass-squared
 matrix in the model file, see \App~\ref{app:modelfile}, and providing explicit
 terms for the ``$\epsilon$ scalars''). So far, Lagrangian parameters as
 appearing in the model files and as printed by the \spheno produced by \sarah 4
 are consistent with the renormalization conditions as specified in the appendix
 of \cite{Hirsch:2012kv}. One should note though that the \vevs from \vcs are
 given for the Landau gauge by default, and have slightly different values to
 those they have in the Feynman-'t Hooft gauge that is used within \spheno, for
 example. 

In principle, every single degree of freedom should be checked for a \vev, but
 this is often totally impractical, given that about ten degrees of freedom is
 at the limit of what might be considered tolerable with current processors.
 Thus the user will often want to consider only a subset of scalars as being
 allowed non-zero \vevs. For example, when considering a supersymmetric model,
 one might restrict oneself to possible \vevs only for the third generation, or
 when considering a model specifically engineered for light staus but all other
 sfermions being very heavy, one might only worry about stau \vevs as a first
 check. Hence \vcs should be used bearing in mind the caveat that it will
 {\it not} find vacua with non-zero \vevs for degrees of freedom which are not
 allowed non-zero \vevs in the model file. It is up to the user to decide on the
 best compromise between speed and comprehensiveness by choosing which degrees
 of freedom to use.

Importantly, the user is responsible for ensuring that the model file has
 a tree-level potential which has a discrete number of minima. Mostly this means
 that the user has to identify unphysical phases and hence remove the associated
 degrees of freedom from the imaginary parts of such complex fields. Another way
 that problems can arise is when there are flat directions such as the
 $\tan\beta=1$ direction of the MSSM with unbroken supersymmetry, even though
 the degeneracy of these directions may be lifted by loop corrections.

\section{Using \vevacious}
\label{sec:automation}

\vevacious needs  at least two input files:
the \textit{model file} and the \textit{parameter file}.
The model file  contains the information about the physical setup.
 This file is  most easily generated by \sarah, as explained
 in \Sec~\ref{sec:sarah_input}. Should the user
 intend to modify  this file by hand, we give more
 information about the format in \App~\ref{app:modelfile}.
The parameter file should be in the \slha format, extended within the
 spirit of the format, as required in general for extended models, as described
 in \Sec~\ref{sec:slha}.
Finally, there is the option of supplying an initialization file in \xml to save
 giving several command-line arguments, and this file is described along with
 these arguments in \Sec~\ref{sec:options}.

\subsection{Preparing the input file for \sarah}
\label{sec:sarah_input}
\sarah \cite{Staub:2008uz, Staub:2009bi, Staub:2010jh,Staub:2012pb} is a tool to
 derive many analytical properties of a particle physics model, like mass
 matrices, tadpole equations, vertices and renormalization group equations,
 from a very short user input. This information can be used, for instance, to
 write model files for several matrix generators or source code for \spheno
 \cite{Porod:2003um, Porod:2011nf}. While previous versions of \sarah were
 optimized for supersymmetric models but supported also to some extent
 non-supersymmetric models, \sarah 4 will provide a simplified input and new
 features also for non-supersymmetric models \cite{Staub:2013tta}. In
 addition, \sarah 4 supports also the output of input files for \vevacious. To
 get this file, run in Mathematica
\begin{lstlisting}
<</path/to/SARAH/SARAH.m;
Start["MyModel"];
MakeVevacious[Options];
\end{lstlisting}
The possible options are:
\begin{itemize}
\item {\tt ComplexParameters}, Value: list of parameters, Default: \{\}: \\
 By default, all parameters are assumed to be real when writing the \vevacious
 input files. However, the user can define those parameters which should be
 treated as complex. 
 \item {\tt IgnoreParameters}, Value: list of parameters, Default: \{\}: \\
 The user can define a list of parameters which should be set to zero when
 writing the \vevacious input.
 \item {\tt OutputFile}, Value: String, Default {\tt MyModel.vin}, where
 {\tt MyModel} here is the same name as is given in {\tt Start["MyModel"];}
 above: \\
 The name used for the output file. 
 \end{itemize}
The first two options allow one to treat parameters differently in the
 \vevacious output as defined in the \sarah model file.
 It may be in the user's interest to try to speed up the evaluation by taking
 out those parameters which on physical grounds play only a subdominant role,
 but the gain by doing so has yet to be quantified.

\paragraph*{Example: MSSM with stau \vevs}
Here we discuss briefly the main steps to prepare an MSSM version including stau
 \vevs. For a more general discussion of the format of the \sarah model files,
 we refer the interested reader to \cite{Staub:2011dp,Staub:2012pb}. In general,
 three changes are always necessary to include new \vevs in a model. 
\begin{enumerate}
 \item Defining the particles which can get a \vev 
 \begin{lstlisting}
  DEFINITION[EWSB][VEVs]= 
  {{SHd0, {vdR, 1/Sqrt[2]}, {sigmad, I/Sqrt[2]},{phid,1/Sqrt[2]}},
   {SHu0, {vuR, 1/Sqrt[2]}, {sigmau, I/Sqrt[2]},{phiu,1/Sqrt[2]}},
   {SeL,  {vLR[3], 1/Sqrt[2]}, {vLI[3], I/Sqrt[2]}, 
                                   {sigmaL, I/Sqrt[2]},{phiL,1/Sqrt[2]}},
   {SeR,  {vER[3], 1/Sqrt[2]}, {vEI[3], I/Sqrt[2]}, 
                                   {sigmaR, I/Sqrt[2]},{phiR,1/Sqrt[2]}},
   {SHdm, {0, 0}, {sigmaM, I/Sqrt[2]},{phiM,1/Sqrt[2]}},
   {SHup, {0, 0}, {sigmaP, I/Sqrt[2]},{phiP,1/Sqrt[2]}},
   {SvL,  {0, 0}, {sigmaV, I/Sqrt[2]},{phiV,1/Sqrt[2]}}
};
\end{lstlisting}
where it is important that the \vevs have names that are at least two
 characters long.
The first two lines are the standard decomposition of the complex Higgs
 scalar $H_i^0 \to \frac{1}{\sqrt{2}}(v_i + i \sigma_i + \phi_i)$ and are the
 same as in the charge-conserving MSSM. The third and fourth line define the
 decomposition of the three generations of left- and right-handed charged
 sleptons. The last three lines define the decomposition of the charged Higgs
 fields and the sneutrinos into CP-even and -odd eigenstates. This is necessary
 for the adjacent mixing, see below.  \\
There are two new features in \sarah 4 which are shown here: (i) it is possible
 to give \vevs just to specific generations of a field, such as in this
 example we only use the third one ({\tt vLR[3]}) -- in the
 same way, one can allow for smuon and stau \vevs using {\tt vLR[2,3]};
 (ii) \vevs can have real and imaginary parts. Until now,
 complex \vevs in \sarah had been defined by an absolute value and
 a phase ($v e^{i \phi}$); however, it is easier to handle within \vevacious in
 the form $v_R + i v_I$. 

\item Changing the rotation of the vector bosons:
\begin{lstlisting}
DEFINITION[EWSB][GaugeSector] =
{ 
  {{VB,VWB[1],VWB[2],VWB[3]},{VB1,VB2,VB3,VB4},ZZ},
  {{fWB[1],fWB[2],fWB[3]},{fWm,fWp,fW0},ZfW} }; 
\end{lstlisting}
With non-zero stau \vevs the photon won't be massless any more but
 will mix with the massive
 gauge bosons. In general, there can be a mixing between the $B$ gauge
 boson ({\tt VB}) and the three $W$ gauge bosons ({\tt VWB[i]}) to four
 mass eigenstates
 ({\tt VB1} \dots {\tt VB4}). The mixing matrix is called {\tt ZZ}. 
\item Changing the rotation of matter fields:
\begin{lstlisting}
DEFINITION[EWSB][MatterSector]= 
{    ...
     {{phid, phiu,phiM,phiP,phiV,phiL,phiR}, {hh, ZH}},
     {{sigmad, sigmau,sigmaM,sigmaP,sigmaV,sigmaL,sigmaR}, {Ah, ZA}},
     {{fB, fW0, FHd0, FHu0, FvL, FeL, conj[FeR],fWm, FHdm, fWp, FHup}, {L0, ZN}}, 
     ...        };  
\end{lstlisting}
The stau \vevs generate new bilinear terms in the scalar potential which
 trigger a mixing between the neutral and charged Higgs fields, the charged
 sleptons and the sneutrinos. Note that even if we had used above complex stau
 \vevs to present the new syntax, we don't take the potential mixing between
 CP-even and -odd eigenstates into account here. To incorporate this, the new
 basis would read
\begin{lstlisting}
DEFINITION[EWSB][MatterSector]= 
{    ...
     {{phid, phiu,phiM,phiP,phiV,phiL,phiR, 
           sigmad, sigmau,sigmaM,sigmaP,sigmaV,sigmaL,sigmaR}, {hh, ZH}},
     ...        };  
\end{lstlisting}
Also, stau \vevs lead to a mixing of all the uncolored fermions, which are now
 also Majorana in nature. In this model file, they are now all labeled as
 {\tt L0} with mixing matrix {\tt ZN}. The spinor sector also has to be
 adjusted:
\begin{lstlisting}
DEFINITION[EWSB][DiracSpinors]={
 Fd ->{  FDL, conj[FDR]},
 Fu ->{  FUL, conj[FUR]},
 Chi ->{ L0, conj[L0]},
 Glu ->{ fG, conj[fG]}
};
\end{lstlisting}
\end{enumerate}

\subsection{Preparing the \slha data}
\label{sec:slha}
To check for the global minimum in a given model for a specific parameter point
 all necessary numerical values of parameters have to be provided via an \slha
 spectrum file. The conventions of this file have to be, of course, identical
 to the ones used for preparing the  \vcs model
 file.

The \slha (1 \cite{Skands:2003cj} and 2 \cite{Allanach:2008qq}) conventions
 are only concerned with the MSSM and the NMSSM, but do specify that
 extra \blocks within the same format should be acceptable within \slha
 files, and should be ignored by programs that do not recognize them. \vcs is
 intended to be used for many other models, so accepts any \blocks that
 are mentioned in its model file and looks for them in the given \slha file. In
 this sense, the user is free to define \blocks as long as the names are
 unbroken strings of alphanumeric characters
 (\eg~\texttt{BLOCK THISISAVALIDNAME} or \texttt{BLOCK MY\_BLOCK\_0123}
\footnote{The \slha papers \cite{Skands:2003cj, Allanach:2008qq} do not
 specify whether \block names should be case-sensitive (so that {\tt HMIX} and
 {\tt Hmix} would be considered equivalent for example) and many spectrum
 generators have already adopted different case conventions for their output,
 so \vcs reads in \block names as case-insensitive.}). However, we strongly advise
 against redefining those \blocks specified by \cite{Skands:2003cj} and
 \cite{Allanach:2008qq}, or any of their elements, to have meanings
 other than those given in \cite{Skands:2003cj} and \cite{Allanach:2008qq}.

With this in mind, two sets of pre-generated model files for the MSSM
 (each file within a set allowing for different scalars to have non-zero \vevs)
 are provided: one set being that produced by \sarah 4, which assumes a certain
 extension of the \slha \texttt{BLOCK HMIX}, the other restricted to quantities
 completely specified by the \slha 1 and 2 conventions. The extensions of
 \texttt{HMIX} are three additional elements: \texttt{101}, providing the value
 of $m_{3}^{2}$ (often written as $B_{\mu}$) directly, and \texttt{102} and
 \texttt{103} providing the values of $v_{d}$ and $v_{u}$ respectively. All
 three can be derived from elements \texttt{2}, \texttt{3}, and \texttt{4} of
 \texttt{HMIX}, but are much more suited to conversion between renormalization
 schemes and different gauges (as $v_{d}$ and $v_{u}$, and thus $\tan\beta$, are
 gauge-dependent quantities, with values which also depend on the
 renormalization conditions).

\subsubsection{Scale dependence in the \slha file}
\label{sec:slha_scales}

\vcs is a tool for finding minima for a one-loop effective potential
 evaluated at a single scale, and thus it is important that the Lagrangian
 parameters are provided consistently at this scale. The scale is also required
 to be given explicitly. Since the \slha convention specifies that running
 parameters are given in \blocks each with their own scale, at first glance this
 may seem problematic. Even worse, the format allows for multiple instances of
 the same \block, each with its own scale. However, the default behaviour of
 \spheno, \softsusy, \suspect, and \isajet when writing \slha output
 is to give all running parameters consistently at a single scale. This is the
 behaviour that \vcs assumes.

The explicit value of the scale $Q$ used in $V^{\text{mass}}$ in
 \eq~(\ref{eq:potential_loop_corrections}) is that given by the
 \block \texttt{GAUGE}. All other \blocks are assumed to be at this same scale
 Although it is not the default behaviour of any of the popular spectrum
 generators to give the same \blocks at different scales, if \vcs finds multiple
 instances of the same \block, the \block with the lowest scale is used and the
 others ignored. In addition, \vcs performs a consistency check that all the
 \blocks used have the same scale, aborting the calculation if not.

\subsubsection{\slha expression of parameters at different loop orders}
\label{sec:tree_or_loop_slha}

The output \blocks enumerated in the \slha papers are specified to be in
 the \drbarp renormalization scheme\footnote{The \drbarp scheme is just
 called the \DRbar scheme in \cite{Skands:2003cj, Allanach:2008qq}, however.},
 but some users may prefer a different
 renormalization scheme. The \slha does not insist on private \blocks adhering
 to the same standards of those explicitly part of the accord, so \vcs allows
 for a certain pattern of private \blocks to give values for a different
 renormalization scheme. (Again, we strongly advise against using the \blocks
 explicitly mentioned in \cite{Skands:2003cj, Allanach:2008qq} to convey values
 that do not adhere to the definitions in
 \cite{Skands:2003cj, Allanach:2008qq}.) The additional renormalization schemes
 that \vcs allows are those where the finite parts of Lagrangian parameters are
 themselves apportioned into loop expansions, \eg~$\mu + \delta \mu$, where
 $\delta \mu$ is considered to be a parameter already of at least one order
 higher than $\mu$.

To allow for different renormalization conditions of this type, \vcs first
 looks for extra (``private'') \slha~\blocks that specify particular loop
 orders. Since \vcs deals with one-loop effective potentials, it has two
 categories of parameters: ``tree-level'' and ``one-loop''. When writing the
 minimization conditions for the tree-level potential, it uses exclusively the
 ``tree-level'' values. When writing the full one-loop effective potential, it
 uses both sets appropriately to avoid including spurious two-loop terms. With
 reference to \eqs~(\ref{eq:tree_level_potential}),
 (\ref{eq:potential_in_parts}), and (\ref{eq:potential_loop_corrections}),
 \vcs writes combines the sum $V^{\text{tree}} + V^{\text{counter}}$ by
 inserting the ``one-loop'' parameter values into $V^{\text{tree}}$
 as part of $V^{\text{1-loop}}$. (The tree-level potential function that is
 also written automatically for convenience, as mentioned in
 \Sec~\ref{subsec:features}, uses the ``tree-level'' values, of course.) The
 term $V^{\text{mass}}$ is already a loop correction, so ``tree-level'' values
 are used in the \MM functions. If only a single value for any parameter is
 given, it is assumed to be in a scheme where it has a single value which is to
 be used in all parts of the effective potential.

As an example, within the renormalization used by \spheno {\tt 3.1.12}, at
 the point SPS1a${}'$, $\mu$ has the value $374.9 \gev$ at tree level, and
 $394.4 \gev$ at one loop. \vcs inserts the value $374.9$ for $\mu$ in the
 minimization conditions (as the units are assumed to be in \gev, as per the
 \slha standard), and also into the ``mass-squared'' matrices that are part of
 the evaluation of the $V^{\text{mass}}$ contributions to the one-loop effective
 potential. \vcs inserts the value $394.4$ for $\mu$ in the polynomial part of
 the potential, accounting for the contributions of both $V^{\text{tree}}$ and
 $V^{\text{counter}}$ together.

In detail, when inserting a ``tree-level'' value, \vcs looks for an \slha
 \block with the prefix ``\texttt{TREE}'' first, and uses that value as
 its ``tree-level'' value. If there is no \block with that prefix, or if there
 is such a \block but it does not specify the appropriate element, then the
 \block without the prefix is assumed to have the appropriate value. Likewise,
 when inserting a ``one-loop'' value, \blocks with the prefix ``\texttt{LOOP}''
 have priority. Hence for the SPS1a${}'$ example above, \vcs could be given an
 \slha file with $394.4$ as element \texttt{1} of \texttt{HMIX}
\footnote{Technically this is already an abuse of the \block {\tt HMIX},
 since the value entering here is not exactly the value it should have in the
 \drbarp scheme, but the difference is a two-loop order effect.} and $374.9$ as
 element \texttt{1} of \texttt{TREEHMIX}. When \vcs is writing the
 ``tree-level'' value of $\mu$, it would first look for element \texttt{1} of
 \texttt{TREEHMIX}, and since it would find $374.9$, this value would be used,
 and element \texttt{1} of \texttt{HMIX} would not be looked for. When writing
 the ``one-loop'' value, element \texttt{1} of \texttt{LOOPHMIX} would be looked
 for, but not found, and because of this, element \texttt{1} of \texttt{HMIX}
 would then be looked for, and $394.4$ would be found and used. One can
 specify element \texttt{1} of both \texttt{TREEHMIX} and \texttt{LOOPHMIX}, and
 then \vcs would never use element \texttt{1} of \texttt{HMIX}, which could give
 the \drbarp value, for example, without worrying
 that it might mix schemes in the calculation. (If one
 prefers to use other prefixes, both ``\texttt{TREE}'' and ``\texttt{LOOP}''
 can be replaced by other strings in the model file in the
 \texttt{$<$block\_prefixes$>$} element;
 \eg~\texttt{$<$block\_prefixes tree="ZEROLOOP" loop="ONELOOP" $/>$}, so that
 \texttt{ZEROLOOPHMIX} and \texttt{ONELOOPHMIX} would be looked for
 appropriately).

Those aspects are taken into account in the
 \spheno~\cite{Porod:2003um,Porod:2011nf} output of \sarah 4
 \cite{Staub:2013tta}. For this purpose, a new flag has been introduced which
 can be used in the \slha input file
\begin{lstlisting}
Block SPhenoInput 	  # SPheno specific input 
 ...
 530 1. 	  # Use Vevacious conventions
\end{lstlisting}
In that case, the new tree and one-loop level block will be present.

\subsection{Setting up and running \vevacious} 
\label{sec:options}
There are many options that can be passed to \vcs. If values other than the
 defaults are required, they can either be passed by command-line arguments or
 with an initialization file in \xml format. If an option is given by both
 commmand-line argument and in an initialization file, the command-line argument
 takes precedence.

We enumerate the options as they would be as command-line arguments setting
 the options to the defaults. All floating point numbers may be given as
 standard decimals, such as {\tt 0.1234}, or in scientific E notation, such as
 {\tt 1.234E-1} (uppercase `E' or lowercase `e', with or without preceeding `0'
 characters, with or without `+' in the exponent, \eg~{\tt 987} or
 {\tt 9.874E+002} or {\tt 0098.7e001}).

\begin{enumerate}
 \item \texttt{--hom4ps2\_dir=./HOM4PS2/}\\ This is a string giving the path to
 the directory where the \homps executable is.

\item \texttt{--homotopy\_type=1}\\  This is an integer used to decide which
 mode \homps should run in: {\tt 1} is used for polyhedral homotopy and {\tt 2}
 for linear homotopy.

\item \texttt{--imaginary\_tolerance=0.0000001}\\ This is a floating-point
 number giving the tolerance for imaginary parts of \vevs found as solutions to
 the tree-level minimization conditions, since it is possible that a numerical
 precision error could lead to what should be an exact cancellation leaving
 behind a small imaginary part. It is in units of \gev, as the other
 dimensionful values are assumed to be so since that is how they are in the
 \slha standard.

\item \texttt{--model\_file=./MyModel.vin}\\ This is a string giving the name
 of the model file discussed in \Sec~\ref{sec:sarah_input}, including the
 (relative or absolute) directory path.

\item \texttt{--slha\_file=./MyParameters.slha.out}\\ This is a string giving
 the name of the \slha file discussed in \Sec~\ref{sec:slha}, including the
 (relative or absolute) directory path.

\item \texttt{--result\_file=./MyResult.vout}\\ This is a string giving
 the name of the \xml output file, discussed in \Sec~\ref{sec:slha}, to write,
 including the (relative or absolute) directory path.

\item \xmle{saddle\_nudges}\texttt{ 1.0, 5.0, 20.0 }\xmle{/saddle\_nudges}\\
 (Unfortunately this option does not work very well as a command-line argument,
 so instead here we display the \xml element as it should appear in the \xml
 initialization file. No initialization file has to be used, of course, if the
 default \texttt{1.0, 5.0, 20.0} is fine.)
 As discussed in \Sec~\ref{sec:vcs_outline}, \pym may get
 stuck at saddle points, and \vcs creates pairs of nearby points as
 new starting points for \pym in an attempt to get it to roll away to minima.
 Using the default list shown, if \vcs finds that \pym stopped at a saddle
 point, it will create two new starting points displaced by $1.0 \gev$ either
 side of this saddle point. If \pym rolls from either or both of these displaced
 points to new saddle points (or just does not roll as it is still in a region
 that is too flat), \vcs will repeat the process for each new saddle point, but
 this time displacing the new starting points by $5.0 \gev$. \vcs can repeat
 this a third time, using $20.0 \gev$, but after this gives up. Giving a longer
 comma-separated list of floating-point numbers will lead to \vcs performing
 this ``nudging'' as many times as there are elements of the list.

\item \texttt{--max\_saddle\_nudges=3}\\ This is an integer giving the length of
 the list of floating-point numbers of the \texttt{saddle\_nudges} option: if it
 is larger than the length of the list \vcs already has, the list is extended
 with copies of the last element; if it is shorter, the list is truncated after
 the given number of elements.

\item \texttt{--ct\_path=./CosmoTransitions}\\ This is a string giving the path
 to the directory where the \ct files \texttt{pathDeformation.py} and
 \texttt{tunneling1D.py} are.

\item \texttt{--roll\_tolerance=0.1}\\ This is a floating-point number giving a
 tolerance for extrema are identified with each other, since \pym may roll to
 the same minimum from two different starting points, but not stop at exactly
 the same point numerically. If the length of the vector that is the
 \textit{difference} of the two field configurations is less than the tolerance
 multiplied by the length of the longer of the two vectors that are the
 displacements of the two field configurations from the origin, then the two
 field configurations are taken to be the same minimum within errors; \eg~if A
 is $v_{d} = 24.42$, $v_{u} = 245.0$ and B is $v_{d} = 24.39$, $v_{u} = 242.7$,
 the length of A is $246.2140$, the length of B is $243.9225$, so the longer
 length is $246.2140$; the length of their difference is $2.300196$ which is
 less than $0.1 * 246.2140$, so A and B are considered to be the same extremum.
 (This is important for avoiding attempting to calculate a tunneling time from
 a point back to itself.)

\item \texttt{--direct\_time=0.1}\\ This is a floating-point number giving a
 threshold tunneling time as a fraction of the age of the Universe for whether
 the metastability is decided by the upper bound from the fast \ct calculation
 taking a straight line from the false vacuum to the true vacuum as described in
 \ref{sec:vcs_outline}. If the upper bound resulting from this calculation is
 below this number, the input vacuum is considered to be short-lived and no
 refinement in calculating the tunneling time is pursued; \eg~if it is $0.1$,
 and the upper bound on the tunneling time is found to be $10^{-20}$ times the
 age of the Universe, the input vacuum is judged to be short-lived since the
 tunneling time is definitely below $0.1$ times the age of the Universe. If
 the value given for \texttt{direct\_time} is negative, this fast calculation is
 skipped.

\item \texttt{--deformed\_time=0.1}\\ This is a floating-point number giving a
 threshold tunneling time as a fraction of the age of the Universe for whether
 the input vacuum is consider short-lived or long-lived when the full \ct
 calculation is performed. If the tunneling time was calculated to have an upper
 bound \textit{above} the threshold given by the \texttt{direct\_time} option
 (or if the calculation of the upper bound was skipped because
 \texttt{direct\_time} was given a negative value), then \ct is called to
 calculate the bounce action allowing it to deform the path in \vev space to
 find the minimal bounce action. If the tunneling time calculated from this
 bounce action is less than the \texttt{deformed\_time} value, the input vacuum
 is considered short-lived, otherwise it is reported to be long-lived. (In order
 to prevent overflow errors when exponentiating a potentially very large number,
 the bounce action is capped at 1000.) If \texttt{deformed\_time} is set to a
 negative value, this calculation is skipped.
      
\end{enumerate}

As mentioned above, an \xml initialization file can be provided with values
 for these options. By default, \vcs looks for
 \texttt{./VevaciousInitialization.xml} for these options, but a different file
 can be specified with the command-line option\\
 \texttt{--input=/example/MyVevaciousInit.xml} to use
 \texttt{/example/MyVevaciousInit.xml} as the initialization file. Any other
 command-ine arguments take priority over options set in the initialization
 file.

An example \xml initialization file called
 {\tt VevaciousInitialization.xml} is provided with the download, which
 shows how to set each option.

It does not matter what the root element is called (the
 example file provided with the download uses \xmle{Vevacious\_defaults}, but it
 really doesn't matter, as long as it is closed properly). Taking the option
 \texttt{slha\_file} as an example, the body of the \xml element with the name
 \texttt{slha\_file} is used as the value of the option (stripped of leading and
 trailing whitespace). Hence
\begin{lstlisting}
  <slha_file>
  /path/to/Vevacious/MSSM/SPheno.spc.MSSM.SPS1ap
  </slha_file>
\end{lstlisting}
would serve instead of
 \texttt{--slha\_file=/path/to/Vevacious/MSSM/SPheno.spc.MSSM.SPS1ap} being
 passed as a command-line argument. Likewise,
\begin{lstlisting}
  <direct_time> 0.01 </direct_time>
\end{lstlisting}
would set the quick calculation threshold to a more conservative time of a
 hundredth of the age of the Universe.

\subsection{The results of \vevacious}
\label{sec:vcs_results}
When \vevacious is finished it returns the results twice: (i) as separate file
 with the name defined in the initialization file, (ii) attached to the used
 \slha spectrum file. The format of the output file is again in \xml. As an
 example of a point which is reported to be stable, we present the result
 of a run on the CMSSM point SPS1a \cite{Allanach:2002nj} (which is strongly
 excluded by experimental non-observation, but suffices as an example). This
 parameter point was checked with the model file described in
 \Sec~\ref{sec:sarah_input} for the MSSM allowing real \vevs for the neutral
 components of the Higgs doublets ({\tt vdR} and {\tt vuR}) and for the
 staus ({\tt vLR3} and {\tt vER3}), and reads
\begin{lstlisting}
<Vevacious_result>
  <reference version="1.0.7" citation="arXiv:1307.1477 (hep-ph)" />
  <stability> stable </stability>
  <global_minimum   relative_depth="-89467096.8481"
                    vdR="24.2105220258" vER3="0.0" vLR3="0.0" vuR="241.158873762" />
  <input_minimum   relative_depth="-89467096.8481"
                   vdR="24.2105220258" vER3="0.0" vLR3="0.0" vuR="241.158873762" />
  <lifetime  action_calculation="unnecessary" > -1.0 </lifetime>
</Vevacious_result>
\end{lstlisting}
(where some line breaks have been inserted into the elements
 \xmle{global\_minimum} and \xmle{input\_minimum} so that they fit on the page).

The element {\tt <stability>} can have the values \texttt{stable}, if the input
 minimum is the global minimum, \texttt{long-lived} if the lifetime of the input
 minima is longer than the specified limit, or \texttt{short-lived} if the
 tunneling time of the input minimum to the global minimum is shorter than the
 specificed threshold.

The elements \xmle{global\_minimum} and \xmle{input\_minimum}
 contain the numerical values of all \vevs at the global and input
 minima respectively as well as the depth of the potential at this point,
 relative to the (real part of the) value of the one-loop effective
 potential for the field configuration where all fields are zero. The \vevs are
 given in units of \gev, while the potential depths are in units of
 (GeV)${}^{4}$.

Finally, \xmle{lifetime} contains information about the method used for the
 calculation of the lifetime of the input minimum as well as the lifetime  as a
 fraction of the age of the Universe (taken as $10^{41}/\gev$ by default, and
 this can be changed by editing line 436 of the default {\tt Vevacious.py}). If
 the global minimum is the input minimum, $-1$ is returned as lifetime. For a
 point with a global charge-breaking minimum (with the provided example
 {\tt CMSSM\_CCB}), again using the MSSM allowing real Higgs and stau \vevs, the
 output file would look like
\begin{lstlisting}
<Vevacious_result>
  <reference version="1.0.7" citation="arXiv:1307.1477 (hep-ph)" />
  <stability> short-lived </stability>
  <global_minimum   relative_depth="-2.11285984487e+13"
                    vdR="4132.33029884" vER3="5551.67597322"
                    vLR3="5115.06350174" vuR="5241.9876933" />
  <input_minimum   relative_depth="-109122205.646"
                   vdR="6.19344185577" vER3="0.0"
                   vLR3="1.13686838E-010" vuR="241.242512796" />
  <lifetime  action_calculation="direct_path_bound" >
             4.38300042027e-26 </lifetime>
</Vevacious_result>
\end{lstlisting}
 (where some linebreaks have been inserted so that the output fits the width of
 the page). We can see that the numerical minimization did not quite roll
 properly to a zero \vev for ${\tilde{\tau}}_{L}$ at the input minimum, but
 stopped extremely close to it. Here, an upper limit of the lifetime has been
 calculated using a direct path between the input and the global minimum. The
 same information is also written to the \slha file using the new block
 {\tt VEVACIOUSRESULTS}, which has elements given by two integer indices
 followed by a floating-point number and a string of characters.
\begin{lstlisting}
BLOCK VEVACIOUSRESULTS # results from Vevacious version 0.3.0, ...
    0   0    -1.00000000E+000    short-lived    # stability of input
    0   1    +4.38300042E-026    direct_path_bound    # tunneling time in Universe ages ...
    1   0    -1.09122206E+008    relative_depth    # input potential energy density ...
    1   1    +0.00000000E+000    vER3    # input VEV
    1   2    +1.13686838E-010    vLR3    # input VEV
    1   3    +6.19344186E+000    vdR    # input VEV
    1   4    +2.41242513E+002    vuR    # input VEV
    2   0    -2.11285984E+013    relative_depth    # global minimum potential energy ...
    2   1    +5.55167597E+003    vER3    # global minimum VEV
    2   2    +5.11506350E+003    vLR3    # global minimum VEV
    2   3    +4.13233030E+003    vdR    # global minimum VEV
    2   4    +5.24198769E+003    vuR    # global minimum VEV
\end{lstlisting}
The conventions are that the {\tt (0,0)} entry of this block gives the
 information about the stability ({\tt -1} for short-lived, {\tt 0} for
 long-lived, {\tt 1} for stable as the floating-point number, followed by
 the description as the second part of the information for these indices).
 For metastable points, the lifetime is saved in {\tt (0,1)} (capped at 1000),
 with the string following giving the type of calculation ({\tt unnecessary} if
 the input minimum was the true vacuum, {\tt direct\_path\_bound} if the upper
 bound from a direct path in field configuration space was below the threshold,
 or {\tt full\_deformed\_path} if \ct had to calculate the action using path
 deformation). The entries {\tt (1,1)} to {\tt (1,n)} contain the numerical
 values of all $n$ \vevs\ and their names as strings at the input minimum and
 {\tt (2,1)} to {\tt (2,n)} give the values and names of the \vevs at the global
 minimum. Entries {\tt (1,0)} and {\tt (2,0)} are the depths of the input and
 global minimum (followed by the string {\tt relative\_depth}). 

\vcs first deletes any files with the same name as was given as the output
 file, and if \vcs was unable to end properly, no output is produced. In the
 case that there  were problems during the run which did not crash the program,
 \vcs prints warnings, and creates an additional \xml element \xmle{warning} in
 the results file, and also appends these warnings in a \block
 {\tt VEVACIOUSWARNINGS}. Possible warnings are (with \texttt{<...>} standing
 for strings describing field configurations, potential depths, or the string
 given by \pym when it throws an exception)
\begin{itemize}
 \item No tree-level extrema were found. This might happen for example if \homps
 did not find any real solutions (recalling that complex fields have already
 been written as pairs of real scalars) because it was given a system with
 continuous degenerate solutions -- however, it is not necessarily the case that
 this is indicative of this problem, and also it is not necessarily guaranteed
 to result from such problematic systems.
 \begin{lstlisting}
 No tree-level extrema were found.
 \end{lstlisting}

 \item \pym threw exceptions:
 \begin{lstlisting}
 PyMinuit had problems starting at <...> [minuit.MinuitError: <...>]. PyMinuit
 stopped at <...> with relative depth <...> at one-loop level and <...> at tree
 level. Minuit's estimate of how much deeper it should go is <...>.
 \end{lstlisting}

 \item \pym got stuck at saddle points with very shallow descending or possibly
 flat directions:
 \begin{lstlisting}e
 <N> extremum/a with at least one descending or flat direction remained after
 all nudging: <...>
 \end{lstlisting}

 \item No one-loop extrema were found. (This shouldn't ever happen, even for a
 potential that is unbounded from below, as by default, \pym is restricted to a
 hypercube of field configurations where no field is allowed a value greater
 than a hundred times the scale $Q$.)
 \begin{lstlisting}
 No one-loop extrema were found.
 \end{lstlisting}

 \item The nearest extremum to the input field configuration is actually a
 saddle point:
 \begin{lstlisting}
 Input VEVs seem to correspond to a saddle point!
 \end{lstlisting}

 \item The nearest extremum to the input field configuration is further away
 than the permitted tolerance:
 \begin{lstlisting}
 PyMinuit rolled quite far from the input VEVs! (from <...> to <...>)
 \end{lstlisting}

 \item The energy barrier between the false and true vacua is thinner than the
 resolution of the tunneling path:
 \begin{lstlisting}
 Energy barrier from input VEVs to global minimum thinner than resolution of
 tunneling path!
 \end{lstlisting}

\end{itemize}

\section{Comparison with existing tools and examples with supersymmetric models}
\label{sec:SUSY_examples}

The major components of \vcs have been tested and used already in the
 literature: \homps, \minuit, and \ct. The innovations of \vcs are the
 automatic preparation and parsing of input and output of the various components
 in a consistent way, optimization of the running time with respect to how
 short-lived a metastable vacuum might be, and the feature that \sarah 4 can
 automatically generate \vcs model files for any new model that can be
 implemented in \sarah.

Very few tools are currently available to do the same job as \vcs. One can
 of course implement the minimization conditions of a tree-level potential in
 \homps or other implementations of the homotopy continuation method, or any
 implementation of the Gr{\"{o}}bner basis method, on a case-by-case basis.

To our knowledge, there is only one publicly-available program that
 purports to find the global minimum of a potential of a quantum field theory:
 {\tt ScannerS} \cite{Coimbra:2013qq}. However, at the time of writing, the
 routines to actually find the global minimum are still under development and
 are not available.

The model files generated by \sarah use the internal expressions that have
 already been cross-checked in \cite{Staub:2010jh,Staub:2011dp}. In addition,
 the potential was always found to have a local minimum at the input field
 configuration, as expected, for a wide range of test points. Furthermore, using
 the model file {\tt SARAH-SPhenoNMSSM\_JustNormalHiggsAndSingletVevs.vin}
 described below and with the modification to \vcs to use only a tree-level
 analysis as described in \Sec~\ref{subsec:features}, we confirmed the results
 of \cite{Maniatis:2006jd}.

Several example model and parameter files are provided with the download of
 \vcs. Three model files are given for the MSSM with different allowed non-zero
 \vevs:
\begin{itemize}
\item {\tt SARAH-SPhenoMSSM\_JustNormalHiggsVevs.vin} with only the normal real
 neutral Higgs \vevs allowed;
\item {\tt SARAH-SPhenoMSSM\_RealHiggsAndStauVevs.vin} with real \vevs allowed
 for the neutral Higgs components and for the staus;
\item {\tt SARAH-SPhenoMSSM\_RealHiggsAndStauAndStopVevs.vin} with real \vevs
 allowed for the neutral Higgs components, for the staus, and for the stops.
\end{itemize}
These model files were generated automatically with \sarah 4. They assume
 that the \slha parameter file will use the standard \sarah-generated
 {\tt SPhenoMSSM} (\spheno using the {\tt MSSM} \sarah model file) output,
 which uses the \slha 2 flavor violation conventions, \ie~that the \blocks\
 {\tt TE}, {\tt TD}, {\tt TU}, {\tt MSQ2}, \etc, are present, and also the
 extra {\tt HMIX} parameters {\tt 101}, {\tt 102}, and {\tt 103}, that
 {\tt SPhenoMSSM} prints out. There are no tadpoles for the first two
 generations of sfermions, which is strictly inconsistent with non-zero \vevs
 for the third generation along with the Higgs doublets in the presence of
 non-zero off-diagonal Yukawa and trilinear soft SUSY-breaking terms. However,
 the assumption is that any point with \eg~non-zero stop \vevs has such small
 sup and scharm \vevs that the stability of the input vacuum can be judged by
 comparison to the minimum in the zero-sup-and-scharm-\vev plane nearest the
 true global minimum.

Three variants which use only \slha2-specified \blocks are also present:\\
 {\tt pure\_SLHA2\_MSSM\_JustNormalHiggsVevs.vin},\\
 {\tt pure\_SLHA2\_MSSM\_RealHiggsAndStauVevs.vin}, and\\
 {\tt pure\_SLHA2\_MSSM\_RealHiggsAndStauAndStopVevs.vin}.\\
These model files also assume that the \slha parameter file will use the
 \slha2 flavor violation conventions, \ie~that the \blocks\ {\tt TE}, {\tt TD},
 {\tt TU}, {\tt MSQ2}, \etc, are present, but do not require the extra
 {\tt HMIX} parameters of the {\tt SARAH-SPheno} versions.

Three variants which use only \slha1-specified \blocks are also present:\\
 {\tt pure\_SLHA1\_MSSM\_JustNormalHiggsVevs.vin},\\
 {\tt pure\_SLHA1\_MSSM\_RealHiggsAndStauVevs.vin}, and\\
 {\tt pure\_SLHA1\_MSSM\_RealHiggsAndStauAndStopVevs.vin}.\\
These model files assume that the \slha parameter file will use the \slha1
 conventions without flavor violation, \ie~that the \blocks\ {\tt AE}, {\tt AD},
 {\tt AU}, \etc, are present, that the sfermion soft SUSY-breaking sfermion
 mass-squared parameters are given by the {\tt MSOFT} \block instead of in
 {\tt MSQ2} \etc, and also do not require the extra {\tt HMIX} parameters of the
 {\tt SARAH-SPheno} versions.

Three model files are also provided for a constrained version of the NMSSM
 (where the dimensionful superpotential parameters $\mu, {\mu}'$, and
 ${\xi}_{F}$, and the soft SUSY-breaking parameters
 $m_{3}^{2}, m_{S}^{\prime2}$, and ${\xi}_{S}$, in the notation of
 \cite{Allanach:2008qq}, are set to zero) with different allowed non-zero \vevs:
\begin{itemize}
\item {\tt SARAH-SPhenoNMSSM\_JustNormalHiggsAndSingletVevs.vin} with only the
 normal real neutral Higgs and singlet \vevs allowed;
\item {\tt SARAH-SPhenoNMSSM\_RealHiggsAndSingletAndStauVevs.vin} with real
 \vevs allowed for the neutral Higgs components, for the singlet, and for the
 staus;
\item {\tt SARAH-SPhenoNMSSM\_RealHiggsAndSingletAndStauAndStopVevs.vin} with
 real \vevs allowed for the neutral Higgs components, for the singlet, for the
 staus, and for the stops.
\end{itemize}
Again, the same flavor issues as the MSSM model files have apply, and
 again, three variants which use only \slha2-specified \blocks are also
 present:\\
 {\tt pure\_SLHA2\_NMSSM\_JustNormalHiggsAndSingletVevs.vin},\\
 {\tt pure\_SLHA2\_NMSSM\_RealHiggsAndSingletAndStauVevs.vin}, and\\
 {\tt pure\_SLHA2\_NMSSM\_RealHiggsAndSingletAndStauAndStopVevs.vin}.
(There are no \slha1 variants, as the NMSSM was not specified in the \slha1
 conventions. It is important to note though that these \slha2 model files still
 also require the flavor violation conventions, \ie~that {\tt TE}, {\tt TD},
 {\tt TU}, {\tt MSQ2}, \etc, are present, even though an NMSSM \slha2 file
 without flavor violation, using {\tt AE} instead of {\tt TE} \etc, is still a
 valid \slha2 file.)

In addition, several example parameter points have been provided to
 demonstrate the existence of metastable points:
\begin{itemize}
\item {\tt SPS1a}, the CMSSM point SPS1a from \cite{Allanach:2002nj} (which is
 stable);
\item {\tt CMSSM\_CCB}, which corresponds to the CMSSM best-fit point
 including LHC and $m_{h} = 126 \gev$ constraints from \cite{Bechtle:2012zk}
 (which has a global charge- and color-breaking minimum);
\item {\tt NUHM1\_CCB}, which corresponds to the NUHM1 best-fit point (``low'')
 from \cite{Buchmueller:2012hv} (which also has a global charge- and
 color-breaking minimum);
\item {\tt CNMSSM\_wrong\_neutral}, which corresponds to
 benchmark point P1 from \cite{Djouadi:2008uw} (which also has a neutral global
 minimum which however is not the input minimum).
\end{itemize}
These parameter files are given in the \slha output of \spheno as
 {\tt SPS1a.slha.out} and so on, with example output as {\tt SPS1a.vout} and so
 on. (The \slha input files are {\tt SPS1a.slha.in} and so on.)

We note that the CCB vacua found for {\tt CMSSM\_CCB}, for example, have \vevs
 of the order of five times the renormalization scale (shown in
 \Sec~\ref{sec:vcs_results}). This should not cause much concern, as
 $\ln( 5^{2} ) / ( 4 \pi ) \simeq 0.256$ so the one-loop effective potential
 should still be reasonable around this minimum. However, even if one restricts
 the \vevs to be less than twice the scale, as described in
 \Sec~\ref{subsec:limitations}, \vcs still finds a tunneling time (upper bound)
 of $10^{-6}$ times the age of the known Universe to a CCB configuration at the
 edge of the bounding hypercube.

The MSSM model and parameter files are in the {\tt MSSM} subdirectory and
 those of the NMSSM are in the {\tt NMSSM} subdirectory.

\section{Conclusion}
\label{sec:conclusion}

Several extensions of the Standard Model contain additional scalar states.
 Usually one engineers the model such that one obtains a phenomenologically
 acceptable vacuum with the desired breaking of
 $SU(2)_L\times U(1)_Y$ to $U(1)_{EM}$. However, in general it is not checked if
 the minimum obtained is the global minimum, as this in general quite an
 involved task already at tree level, where hardly any analytical conditions can
 be given. Often loop corrections become important too. 

To tackle this problem, we have presented the program \vcs as a tool to quickly
 evaluate one-loop effective potentials of a given model. It finds all extrema
 at tree level, which allows for a first check for undesired minima. Starting
 from these extrema, it calculates the one-loop effective potential to obtain a
 more reliable result. This is important as loop contributions can potentially
 change the nature of an extremum. In the case that the original minimum turns
 out to be merely a local minimum rather than the global minimum, the
 possibility is given to evaluate the tunneling time.  As test cases we have
 considered supersymmetric models, but the program can be used for a general
 model provided that the tree-level potential is polynomial in the scalar
 fields.

\section*{Acknowledgments}
The authors would like to thank Max Wainwright for advice on \ct.
 This work has been supported by the DFG project No.~PO-1337/2-1 and
 the research training group GRK 1147.


\appendix
\section{Installation and pre-requisites}
\lstset{basicstyle=\small, frame=none}
To fully evaluate a parameter point with \vcs
and the connected tools, you need a Linux or MacOS system
with \dots
\begin{itemize}
 \item  \dots a \cpp compiler such as {\tt gcc}
 \item  \dots the \py environment, at least version {\tt 2.7.1}
 \item  \dots the {\tt python-dev} headers ({\tt Python.h})
 for {\tt gcc} or the equivalent -- {\it this is important!} ({\tt MacPorts}
 for {\tt OSX} automatically installs {\tt Python.h} as part of the installation
 of \py\ {\tt 2.7})
\end{itemize}
If you want to create new input files using \sarah, you need at least
 {\tt Mathematica 7}. 
To compile \spheno to create the numerical input for \vevacious, a 
{\tt Fortran} compiler such as {\tt gfortran} or {\tt ifort} is needed. \\

\vevacious makes use of  several public tools. We give
 here only a brief introduction to the installation of these tools but refer
 to the corresponding references and authors for more information.

We assume in this descrption that all codes are downloaded and extracted in the
 same directory. The placeholder for the path to this directoy is called
 {\tt \$VPATH} in the following. 
\begin{enumerate}
 \item \homps
 \begin{enumerate}
 \item Download \homps from \\ \hspace*{-9mm}
 {\tt
 http://www.math.nsysu.edu.tw/$\sim$leetsung/works/HOM4PS\_soft\_files/HOM4PS\_Linux.htm}
  \item Extract the tar-file
  \begin{lstlisting}
  > tar -xf HOM4PS2_64-bit.tar.gz 
  \end{lstlisting}
  which should create {\tt \$VPATH/HOM4PS2/}.
  \item No compilation is necessary\, but now would be a good point to check
        that \homps works with one of its examples.
  \end{enumerate}
  \item \minuit and \pym
  Detailed instructions about the installation and compilation of \pym and
  \minuit are given at
  {\tt https://code.google.com/p/pyminuit/wiki/HowToInstall}
  We summarize only briefly the main steps
  \begin{enumerate}
   \item Download \minuit from {\tt http://code.google.com/p/pyminuit/}.\\
   Choose {\tt Minuit-1\_7\_9-patch1.tar.gz}.
   \item Extract the tarball
   \begin{lstlisting}
    > tar -xf Minuit-1_7_9-patch1.tar.gz
   \end{lstlisting}
  which should create {\tt \$VPATH/Minuit-1\_7\_9/}.
  \item Configure and compile \minuit 
    \begin{lstlisting}
     > cd Minuit-1_7_9
     > ./configure --prefix=$VPATH/Min/
     > make
     > make install
    \end{lstlisting}
  which should create {\tt \$VPATH/Min/}.
  \item Download {\tt pyminuit-1.2.1.tar.gz} or a more recent version from\\
        {\tt http://code.google.com/p/pyminuit/}
  \item Extract \pym
  \begin{lstlisting}
   > tar -xf pyminuit-1.2.1.tar.gz
  \end{lstlisting}
  which should create {\tt \$VPATH/pyminuit/}.
  \item Run the setup of \pym stating the location of \minuit\ {\it where the}
 {\tt .o} {\it files are}, so {\tt \$VPATH/Minuit-1\_7\_9/} rather than
 {\tt \$VPATH/Min/}
  \begin{lstlisting}
   > cd pyminuit-1.2.1
   > python setup.py install --home=$VPATH/pym/
                             --with-minuit=$VPATH/Minuit-1_7_9/
  \end{lstlisting}
  which should create {\tt \$VPATH/pym/} (with the second command broken over
 two lines to fit on the page).
  \end{enumerate}
 \item Make sure that you have properly exported {\tt PYTHONPATH} (which should
 include the path to {\tt minuit.so}, in {\tt \$VPATH/pym/lib/python/} in our
 example) and {\tt LD\_LIBRARY\_PATH} (which should
 include the path to {\tt liblcg\_Minuit.a}, in {\tt \$VPATH/Min/lib/} in our
 example). Now would be a good time to check that \pym works by running the
 following test program
  \begin{lstlisting}
   >>> import minuit
   >>> def f( x, y ): return ( ( x**2 + y**2 )**2 - 0.5 * x**2 )
   >>> m = minuit.Minuit( f )
   >>> m.values = { 'x': 0.1, 'y': 0.2 }
   >>> m.migrad()
   >>> m.values, m.fval
  \end{lstlisting}
 \item \ct
 \begin{enumerate}
  \item Ensure that the \py packages {\tt Numpy} and {\tt SciPy} are installed
        (if not, please use your favorite Internet search engine to find out how
        to install them on your system)
  \item Download {\tt CosmoTransitions\_package\_v1.0.2.zip} or a more recent
        version from {\tt http://chasm.uchicago.edu/cosmotransitions}
  \item unzip the archive
  \begin{lstlisting}
   > unzip CosmoTransitions_package_v1.0.2.zip
  \end{lstlisting}
  which should create {\tt \$VPATH/CosmoTransitions\_package\_v1.0.2/}
 \end{enumerate}
 \item \lhpc
 \begin{enumerate}
   \item Download \lhpc from {\tt http://www.hepforge.org/downloads/lhpc/}
   \item Extract the  files from the compressed tarball
         (change the version number appropriately)
   \begin{lstlisting}
    > tar -xf LHPC-0.8.5.tar.gz
   \end{lstlisting}
   \item Enter the \lhpc directory and compile it 
   \begin{lstlisting}
    > cd LHPC-0.8.5
    > make
   \end{lstlisting}
 \end{enumerate}
 \item \vcs
 \begin{enumerate}
 \item Download the most recent version from\\
       {\tt http://www.hepforge.org/downloads/vevacious/}
 \item Extract the  files from the compressed tarball and
       compile \vcs stating the path to \lhpc
 \begin{lstlisting}
  > tar -xf  Vevacious-1.0.7.tar.gz
  > cd Vevacious-1.0.7
  > make LHPCDIR=$VPATH/LHPC-0.8.5/
 \end{lstlisting}
 Now would be a good time to check that \vcs works. Edit\\
 {\tt \$VPATH/Vevacious-1.0.7/bin/VevaciousInitialization.xml} to correctly set
 all the paths, then run
 \begin{lstlisting}
  > ./bin/Vevacious.exe --input=./bin/VevaciousInitialization.xml
 \end{lstlisting}
 which should produce the result file with the name given in the initialization
 file.
 \end{enumerate}
   For example, in {\tt Debian}-based Linux distribution one can ensure that the
 paths are properly set by adding the following to {\tt .bashrc}:
  \lstset{basicstyle=\scriptsize, frame=none}
  \begin{lstlisting}
   export PYTHONPATH=$VPATH/pym/lib/python/:$PYTHONPATH
   export PYTHONPATH=$VPATH/CosmoTransitions_package_v1.0.2/cosmoTransitions/:$PYTHONPATH
   export LD_LIBRARY_PATH=$VPATH/Min/lib/:$LD_LIBRARY_PATH
  \end{lstlisting}
\lstset{basicstyle=\small, frame=none}
\end{enumerate}
To get the input for a point, you can use a spectrum generator based on \spheno 
 based on the corresponding \sarah output
\begin{enumerate}
 \item \sarah 
 \begin{enumerate}
   \item Download the most recent version of \sarah 4 from
 {\tt http://sarah.hepforge.org}
   \item Extract \sarah and run {\tt Mathematica}
   \begin{lstlisting}
    > tar -xf SARAH4b-0.0.5.tar.gz
    > mathematica
   \end{lstlisting}
   \item Load \sarah, evaluate a model and generate the \spheno output
   \begin{lstlisting}
    <<"$VPATH/SARAH4b-0.0.5/SARAH.m";
    Start["MyModel"];
    MakeSPheno[];
   \end{lstlisting}
  {\tt MyModel} can be for instance {\tt MSSM} or {\tt NMSSM}. 
 \end{enumerate}
 \item \spheno
 \begin{enumerate}
 \item Download \spheno from {\tt http://spheno.hepforge.org}
 \item Extract \spheno 
 \begin{lstlisting}
  > tar -xf SPheno-3.2.3.tar.gz
  \end{lstlisting}
 \item Enter the \spheno directory and create a new subdirectory for the new model
 \begin{lstlisting}
  > cd SPheno-3.2.3
  > mkdir MyModel
 \end{lstlisting}
 \item Copy the source code produced by \sarah to the new sub directory
 \begin{lstlisting}
  > cp $VPATH/SARAH4b-0.0.5/Output/MyModel/EWSB/SPheno/* MyModel/
 \end{lstlisting}
 \item Compile \spheno and the new \spheno module 
 \begin{lstlisting}
  > make Model=MyModel
 \end{lstlisting}
 \end{enumerate}
\end{enumerate}

\section{Model file format}
\label{app:modelfile}
\lstset{basicstyle=\scriptsize, frame=shadowbox}
\vcs is not restricted to a specific model, but any necessary information
 to evaluate the one-loop effective potential in a given model, assuming a
 particular set of \vevs,  must be provided by the user as a model file. This
 file can usually be produced by \sarah as described in
 \Sec~\ref{sec:sarah_input}. 
 However, in the case that the user wants to modify things manually, we give
 here more information about the format. \\

The Lagrangian parameters that will take values given by the \slha file are
 written in the form {\tt SLHA::BLOCKNAME[ENTRY]}, where {\tt ENTRY} is a
 comma-separated list of indices. For example, element {\tt 1} of the {\tt HMIX}
 \block is written as {\tt SLHA::HMIX[1]} and the {\tt 2,3} element of the
 {\tt YD} \block is written as {\tt SLHA::YD[2,3]} (floating point numbers are
 still interpreted as integers, so {\tt SLHA::YD[2.0,3.0]} would also work).
 Any \blocks with no index should use an empty list, \eg~{\tt SLHA::ALPHA[]}.

This file is in \xml and begins/ends with 
\begin{lstlisting}
<Vevacious_stuff> 
...
</Vevacious_stuff> 
\end{lstlisting}
(though the actual name of the root element is not important: one could use
\begin{lstlisting}
<MyModelForVevacious> 
...
</MyModelForVevacious> 
\end{lstlisting}
 for example instead).
The model file must contain the following information:
\begin{enumerate}
 \item {\bf The names of all \vevs which can be present}
\begin{lstlisting}
<input_vevs vdR="SLHA::HMIX[102]" vuR= "SLHA::HMIX[103]" vLR3 ="0" vER3 = "0" >
 <taken_positive> vdR, vE3 </taken_positive>
</input_vevs>  
\end{lstlisting}
The \xml element \xmle{input\_vevs} serves the dual purpose of enumerating the
 allowed non-zero \vevs along with specifying what is considered to be the input
 minimum. In the MSSM example here, the Higgs fields of course are allowed
 \vevs, and the input minimum is specified by {\tt vdR} ($v_{d}$) being provided
 in the \slha file in the \block {\tt HMIX} as entry {\tt 102}, and
 {\tt vuR} ($v_{u}$) by entry {\tt 103}. The model file is also allowing
 non-zero stau \vevs ({\tt vLR3, vER3}), and also at the same time specifying
 that their value at the input minimum is {\tt 0}. (One can actually put valid
 \py code here within the quote marks for the input minimum values, though it is
 not recommended, and surrounding the code with brackets is advised if one
 insists on going through with it.)

 In addition, the is usually some redundancy in the
 sets of \vevs which minimize the tree-level scalar potential, since different
 minima can be related by phase rotations. To reduce the number of redundant
 solutions, it can be explicitly defined that some \vevs have to be positive
 using {\tt <taken\_positive> ... </taken\_positive>}.
\item {\bf The tree-level tadpole equations}
\begin{lstlisting}
<tadpoles> 
{ 
...
(0.5*vL3*SLHA::MSL2[1., 3.]) 
+(0.5*vL3*SLHA::MSL2[3., 1.]) 
+(0.7071067811865475*vdR*vE3*SLHA::TE[3., 1.]) 
+(0.5*vdR^2*vL3*SLHA::YE[1., 1.]*SLHA::YE[1., 3.]) 
+(0.5*vdR^2*vL3*SLHA::YE[2., 1.]*SLHA::YE[2., 3.]) 
+(-0.7071067811865475*vE3*vuR*SLHA::HMIX[1.]*SLHA::YE[3., 1.]) 
+(0.5*vdR^2*vL3*SLHA::YE[3., 1.]*SLHA::YE[3., 3.]) 
+(0.5*vE3^2*vL3*SLHA::YE[3., 1.]*SLHA::YE[3., 3.]) 
; 
...
}
</tadpoles> 
\end{lstlisting}
This block contains a list of entries in the format
 $(t_1; t_2; t_3; \dots t_n;)$. Here, $t_i$ are the tadpole equations of the
 tree-level scalar potential,
 \ie~$t_i = \frac{\partial V}{\partial {\phi}_{i}} = 0$.
 Note that each equation has to end with a semi-colon. In addition, to
 circumvent problems during parsing this file, it is convenient to use for
 each term a separate line and to put it into brackets. Furthermore, to
 associate the different parameters with the numerical values given later on via
 an \slha spectrum file, all parameters but the field configurations have to be
 replaced by their corresponding entries in the \slha file, in the
 {\tt SLHA::BLOCKNAME[ENTRY]} format explained above. For instance, using the
 \slha 2 conventions \cite{Allanach:2008qq}, the hypercharge $g_1$ is replaced
 by {\tt SLHA::GAUGE[1]} and the top Yukawa coupling $Y_t = Y_u^{33}$ by
 {\tt SLHA::YU[3,3]}. 
 \item  {\bf The polynomial part of the scalar potential}
 \begin{lstlisting}
<polynomial_part> 
(0.03125*vdR^4*SLHA::GAUGE[1.]^2) 
  + (-0.125*vdR^2*vE3^2*SLHA::GAUGE[1.]^2) 
  + (0.125*vE3^4*SLHA::GAUGE[1.]^2) 
  + (0.0625*vdR^2*vL3^2*SLHA::GAUGE[1.]^2) 
  + (-0.125*vE3^2*vL3^2*SLHA::GAUGE[1.]^2) 
 ...
</polynomial_part>
 \end{lstlisting}
 This block contains the scalar potential $V({\phi}_i;g_i,Y_i,T_i,\dots,m^2_i)$
 as function of all possible field configurations and the
 other parameters like gauge and Yukawa
 couplings or mass terms. The conventions are similar to those of
 {\tt <tadpoles>}: (i) choose a separate line for each term, (ii) put each
 term into brackets, (iii) replace all parameters but the field configurations
 by their \slha entries. 
 \item {\bf All mass-squared matrices to calculate the full one-loop effective
 potential}
 \begin{lstlisting}
 <mass-squared_matrix 
particle="Sd"  rotationmatrix="ZD"  factor="6" >     
(-0.041666666666666664*vdR^2*SLHA::GAUGE[1.]^2+...), 
...
</mass-squared_matrix> 
 \end{lstlisting}
 To calculate the one-loop effective potential all field-configuration-dependent
 ``masses'' \MM have to be specified. This happens by using for each
 mass-squared matrix the \xml element {\tt <mass-squared\_matrix ...>}.
 Two attributes must be given: an overall constant ({\tt factor}) which takes
 into account the degrees of freedom of the particle, including the
 spin $s_{n}$ from \eq~\ref{eq:potential_loop_corrections}, and also saves
 reproducing identical matrices due to color factors or pairs of charge
 conjugates, for example. This factor is given by
 $r\cdot c_{F} \cdot (-1)^s (2s+1)$, where $r=1$ holds
 for real bosons or Majorana fermions, and $r=2$ for complex bosons and Dirac
 fermions, and $c_{F}$ is the number of degenerate states (\eg~$6 = 3$
 colors $\times 2$ charge-conjugate states for quarks, if $SU(3)_{c}$ is
 unbroken in the model file). \\
 The body of each block contains all entries of the mass-squared matrix. Here,
 the convention is that each line consists of one element of the mass matrix
 which is placed into brackets and ends with a comma. The order is that first
 all elements of a line are given from left to right,
 before the entries of the next line follow, \ie~for an $n\times n$
 mass-squared matrix, the order is
 $(({\bar{M}}^{2}_{11}), ({\bar{M}}^{2}_{12}),
 .., ({\bar{M}}^{2}_{1n}), ({\bar{M}}^{2}_{21}),
 .., ({\bar{M}}^{2}_{n1}), .. ,({\bar{M}}^{2}_{nn}))$. 
\end{enumerate}
It is, of course, necessary to express the tadpole equations, the potential as
 well as the mass matrices taking into account all field configurations that the
 user wants to check. For instance, to study charge-breaking minima in the MSSM,
 the tadpole equations for the stau \vevs have to be given as input as well as
 all possible terms in potential. Furthermore, the mass matrices must include
 the mixing between the Higgs fields, the sneutrinos and the charged sleptons
 which can be triggered by non-vanishing stau \vevs. If, in addition, color
 conservation should be checked, also stop \vevs have to be included with their
 entire impact. Obviously, preparing this input by hand can easily become a
 cumbersome task. Therefore, we recommend automatic generation of the input
 using the \mathematica package \sarah.

\bibliography{Vevacious_manual.bib}
\bibliographystyle{h-physrev5}

\end{document}